\documentclass[letterpaper,twocolumn,10pt]{article}
\usepackage{usenix-2020-09}

\usepackage{tikz}
\usepackage{amsmath,amssymb,amsfonts}
\usepackage{algorithmic}
\usepackage{graphicx}
\usepackage{textcomp}
\usepackage{xcolor}
\usepackage{fancyhdr}
\newcommand{\myparatight}[1]{\smallskip\noindent{\bf {#1}:}~}
\usepackage{makecell}
\usepackage{multirow}
\usepackage{subfig}
\usepackage[ruled,linesnumbered]{algorithm2e}

\usepackage{bbm}
\usepackage{pdfx}

\newcommand\SHF[1]{{#1}}
\begin{document}
\date{}

\title{\Large \bf PoisonedEncoder: Poisoning the Unlabeled Pre-training Data \\ in Contrastive Learning}

\author{
{\rm Hongbin Liu\quad \quad Jinyuan Jia\quad \quad Neil Zhenqiang Gong}\\
Duke University\\
\{hongbin.liu, jinyuan.jia, neil.gong\}@duke.edu\\
} 

\maketitle

\pagestyle{empty}
\begin{abstract}
Contrastive learning pre-trains an image encoder using a large amount of unlabeled data such that the image encoder can be used as a general-purpose feature extractor for various downstream tasks. 
In this work, we propose \emph{PoisonedEncoder}, a \emph{data poisoning attack} to contrastive learning. In particular, an attacker injects  carefully crafted poisoning inputs into the unlabeled pre-training data, such that the downstream classifiers built based on the poisoned  encoder for multiple  target downstream tasks simultaneously classify attacker-chosen, arbitrary clean inputs as attacker-chosen, arbitrary classes. We formulate our data poisoning attack as a bilevel optimization problem, whose solution is the set of poisoning inputs; and we propose a contrastive-learning-tailored method to approximately solve it. Our evaluation on multiple datasets shows that PoisonedEncoder achieves high attack success rates while maintaining the testing accuracy of the downstream classifiers built upon the poisoned encoder for non-attacker-chosen inputs. We also evaluate five defenses against PoisonedEncoder, including one \emph{pre-processing}, three \emph{in-processing}, and one \emph{post-processing} defenses. Our results show that these defenses can decrease the attack success rate of  PoisonedEncoder, but they also sacrifice the utility of the encoder or require a large clean pre-training dataset. 
\end{abstract}
\section{Introduction}

\begin{figure}[!t]
    \centering
    {\includegraphics[width=0.48\textwidth]{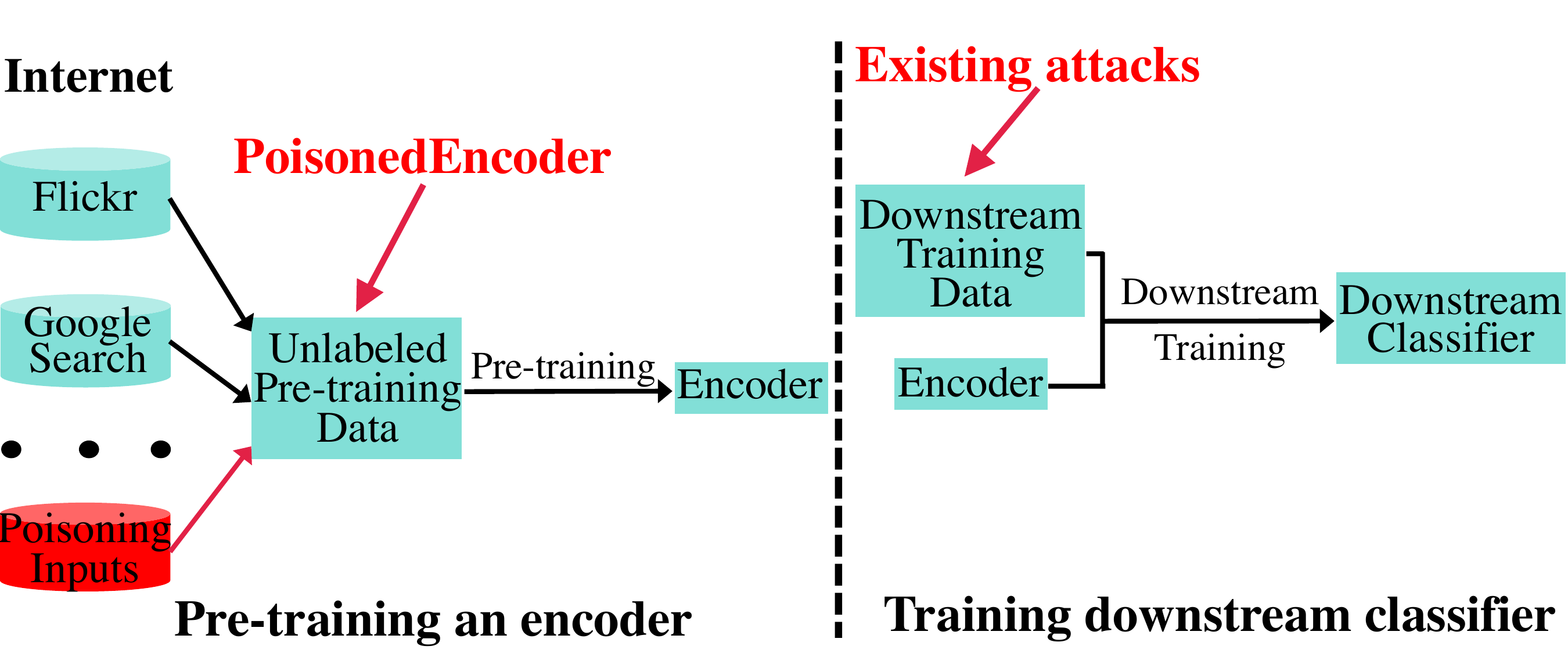}}
    \caption{Our PoisonedEncoder vs. existing data poisoning attacks in the contrastive learning pipeline.}
    \label{fig-comparison}
\end{figure}

Contrastive learning~\cite{he2020momentum, chen2020simple, radford2021learning} is an emerging machine learning paradigm. Specifically,  an \emph{encoder provider} (e.g., Google, OpenAI, and Meta) pre-trains encoders (we focus on image encoders) using a large amount of unlabeled data (called \emph{pre-training data}) automatically collected from the Internet via a crawler; and the image encoders are then used as general-purpose feature extractors for various downstream tasks. The unlabeled data could be images (known as \emph{single-modal contrastive learning})~\cite{chen2020simple,he2020momentum} or (image, text) pairs (known as \emph{multi-modal contrastive learning})~\cite{radford2021learning}. Given an image encoder as a feature extractor and  a small amount of \emph{downstream training data} for a downstream task (e.g., traffic sign recognition, face mask detection, and face recognition), a downstream classifier can be trained using the standard supervised learning or semi-supervised learning.

In this work, we study targeted {data poisoning attack} to contrastive learning. Specifically, an attacker aims to make the downstream classifiers built for multiple downstream tasks (called \emph{target downstream tasks}) misclassify  arbitrary, attacker-chosen clean inputs (called \emph{target inputs}) as arbitrary, attacker-chosen classes (called \emph{target classes}).
For instance, the attacker may desire a target traffic-sign classifier to misclassify a stop sign (target input) as speed limit (target class) and a target face-mask classifier to misclassify a person not wearing mask (target input) as wearing mask (target class). Such attacks pose significant challenges to contrastive learning in safety and security critical applications.

Existing targeted data poisoning attacks mainly focus on supervised learning~\cite{munoz2017towards,suciu2018does,shafahi2018poison,geiping2020witches} and semi-supervised learning~\cite{wang2019attacking,carlini2021poisoning}. When applied to  attack target downstream classifiers in contrastive learning, they aim to tamper with their downstream training data, as illustrated in Figure~\ref{fig-comparison}. As a result, these attacks are not applicable when the downstream training data maintains integrity. For instance, when the downstream training data is proprietary data obtained from a trustworthy source; and since the  downstream training datasets are often small in contrastive learning, they may also be manually cleaned up.
However, the unlabeled pre-training data in contrastive learning is often collected from the Internet and thus is vulnerable to poisoning. For instance, an attacker can publish its poisoning images/inputs on crawler-accessible websites such as social media websites, so they can be collected as a part of the pre-training data by an encoder provider. Therefore, we focus on poisoning the unlabeled pre-training data in this work. A recent work~\cite{carlini2021poisoning_clip} studied targeted data poisoning attack to contrastive learning. However, they focused on multi-modal contrastive learning. In particular, their attack injects poisoning (image, text) pairs into the pre-training data, where the image is a target input and the text includes the target class name, such that the poisoned encoders produce similar feature vectors for the target input and target class. Their attack is not applicable to single-modal contrastive learning because it does not use text. \SHF{Another recent work~\cite{jia2021badencoder} studied backdoor attacks to contrastive learning. However, they compromise the pre-training process, while we poison the pre-training data, which is a more realistic threat model.}

\myparatight{Our work} We propose PoisonedEncoder, the first targeted data poisoning attack to single-modal contrastive learning. PoisonedEncoder injects carefully crafted poisoning inputs into the unlabeled pre-training data such that multiple target downstream classifiers trained based on the poisoned encoder misclassify the target inputs as target classes simultaneously.

The key challenge of PoisonedEncoder is to craft the poisoning inputs to achieve the attack goals. To address the challenge,   we formulate PoisonedEncoder as a bilevel optimization problem, whose solution is the set of poisoning inputs. 
Specifically, the outer optimization problem captures the attack goals on the poisoned encoder, while the inner optimization problem captures that the poisoned encoder is learnt on the poisoned pre-training data. However, it is notoriously challenging to solve bilevel optimization problems. To address the challenge, we propose an approximate solution that is tailored to contrastive learning. 
Specifically, contrastive learning aims to learn an encoder that produces similar feature vectors for two randomly cropped views of an image. Based on this observation, we concatenate a target input and an image (called \emph{reference input}) in the target class either horizontally or vertically to construct a poisoning input. Since two randomly cropped views of a poisoning input may correspond to the target input and reference input, the poisoned encoder may produce similar feature vectors for the target input and reference input. Therefore, a target downstream classifier built upon the poisoned encoder is likely to predict the same class for the target and reference inputs, which is the target class.

We evaluate PoisonedEncoder on multiple datasets in different settings. On one hand, PoisonedEncoder achieves high attack success rate, i.e., the target downstream classifiers built based on the poisoned encoder misclassify a large fraction of target inputs as the target classes. On the other hand, \SHF{PoisonedEncoder} maintains the encoder's utility, i.e., a downstream classifier built based on a clean encoder and that built based on a poisoned encoder for a target/non-target downstream task have similar accuracy for non-target inputs.  \SHF{PoisonedEncoder} can attack multiple target inputs and multiple target downstream tasks simultaneously. Moreover, we extend state-of-the-art targeted data poisoning attacks designed for supervised learning~\cite{geiping2020witches} and semi-supervised learning~\cite{carlini2021poisoning} to poison the pre-training data in contrastive learning. Our results show that PoisonedEncoder achieves higher attack success rate than these extended attacks.

Defenses against data poisoning attacks can be roughly categorized into \emph{pre-processing}, \emph{in-processing}, and \emph{post-processing}. We explore one pre-processing defense (i.e., detecting and removing poisoning inputs before pre-training an encoder), three in-processing defenses (i.e., early stopping of pre-training an encoder and training a downstream classifier, ensemble method, and pre-training without random cropping), and one post-processing defense (i.e., fine-tuning a potentially poisoned encoder using a clean pre-training dataset). Our results show that these defenses can reduce the attack success rate of PoisonedEncoder, but they sacrifice the utility of the encoder or require substantial manual efforts to collect a large clean pre-training dataset. 

Our key contributions can be summarized as follows:
\begin{itemize}
    \item We propose PoisonedEncoder, the first data poisoning attack to single-modal contrastive learning. 
    \item We formulate PoisonedEncoder as a bilevel optimization problem and propose a contrastive-learning-tailored method to solve it. 
    \item We extensively evaluate PoisonedEncoder in different settings and compare it with state-of-the-art attacks extended to contrastive learning. 
    \item We explore five defenses against PoisonedEncoder.
\end{itemize}

\section{Background on Contrastive Learning}
The  pipeline of contrastive learning consists of two stages, i.e., pre-training an encoder and training downstream classifiers. Next, we discuss these two stages in more detail.

\subsection{Pre-training an Encoder}
\begin{figure}[!t]
    \centering
    {\includegraphics[width=0.45\textwidth]{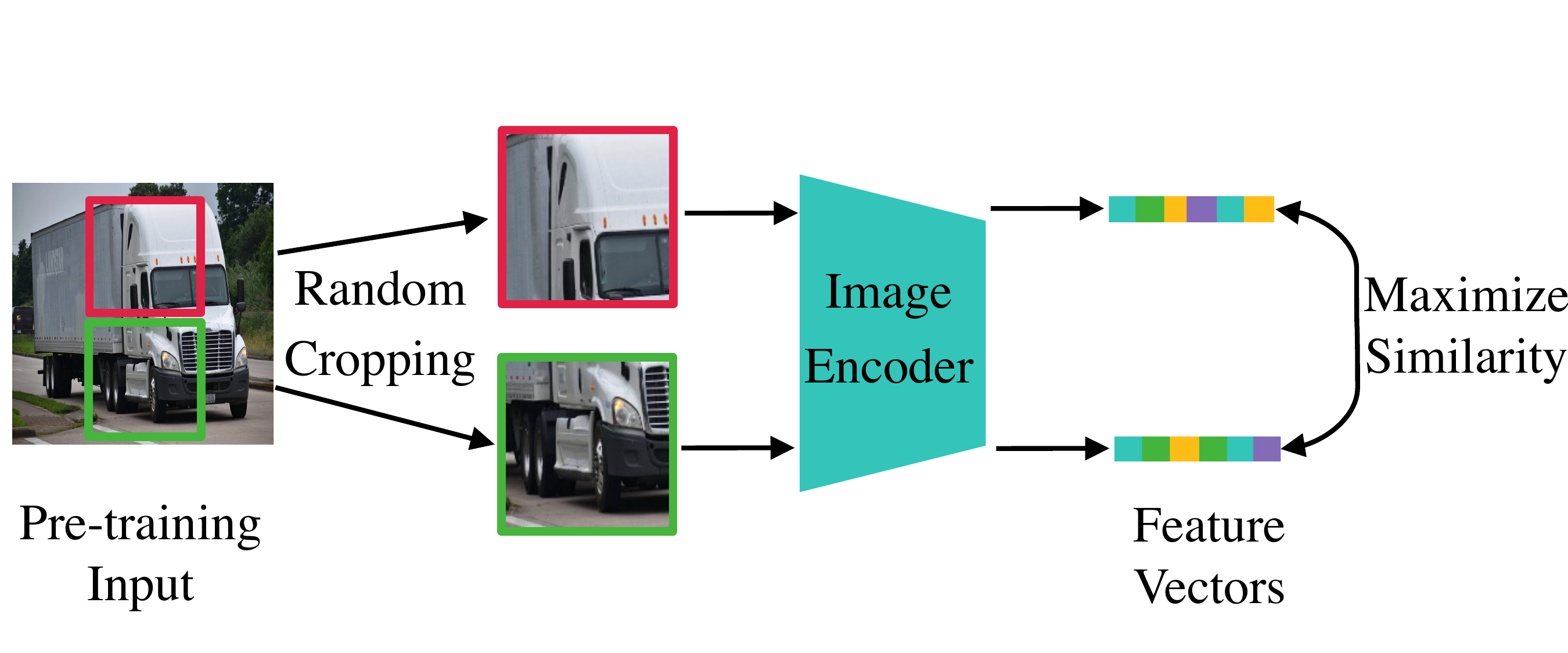}}
    \caption{Illustration of contrastive learning when random cropping is used to generate augmented views of a pre-training input. } 
    \label{fig-random-crop}
\end{figure}

A key module of pre-training an encoder is \emph{random data augmentation}.  
Given a pre-training input, the random data augmentation module generates two augmented views via a series of random augmentation operations, such as random cropping, random horizontal flipping, and randomly converting to grayscale. Two augmented views are named \emph{positive pair} (or \emph{negative pair}) if they are from the same (or different) pre-training input(s). Roughly speaking, the core idea of contrastive learning is to pre-train an encoder that outputs similar (or dissimilar) feature vectors for a positive (or negative) pair.  
In particular, \emph{random cropping}, which first crops a random region with a certain size from an input and then resizes the region to have the same size as the input, is a crucial data augmentation operation for contrastive learning~\cite{chen2020simple,he2020momentum}. 
Figure~\ref{fig-random-crop} illustrates the core idea of contrastive learning when random cropping is used as a data augmentation operation. As we will discuss, our PoisonedEncoder exploits the random cropping operation to poison contrastive learning. Next, we describe two representative contrastive learning algorithms, SimCLR~\cite{chen2020simple} and MoCo~\cite{he2020momentum}. 

\myparatight{SimCLR~\cite{chen2020simple}}
Besides the data augmentation module, two major components of SimCLR are the  encoder $f$ and the projection head $h$. The  encoder $f$ generates a feature vector $f(\mathbf{\mathbf{x}})$ for an input $\mathbf{x}$, while the projection head $h$ performs non-linear mapping for the feature vector $f(\mathbf{x})$ via a multilayer perceptron, which is used to calculate contrastive loss. Given a mini-batch of $K$ pre-training inputs (denoted as $\{\mathbf{x}_1, \mathbf{x}_2, \cdots, \mathbf{x}_K\}$), each of them is transformed into two randomly augmented views via the data augmentation module, producing $2 \cdot K$ augmented inputs: $\{\mathbf{x}_1^{\prime}, \mathbf{x}_2^{\prime}, \cdots, \mathbf{x}_{2 \cdot K}^{\prime}\}$. For simplicity, we use $\mathbf{u}_i$ to denote the projection head's output for the augmented view $\mathbf{x}_i^{\prime}$, i.e., $\mathbf{u}_i=h\left(f\left(\mathbf{x}_i^{\prime}\right)\right)$, where $i=1,2,\cdots,2\cdot K$. In particular, given a positive pair $(\mathbf{x}_i^{\prime},\mathbf{x}_j^{\prime})$, SimCLR defines its contrastive loss as follows:
\begin{align}
\ell_{i, j}=-\log \left( \frac{\exp \left(\operatorname{sim}\left(\mathbf{u}_{i}, \mathbf{u}_{j}\right) / \tau\right)}{\sum_{k=1}^{2 \cdot K} \mathbb{I}(k \neq i) \cdot \exp \left(\operatorname{sim}\left(\mathbf{u}_{i}, \mathbf{u}_{k}\right) / \tau\right)}\right),
\end{align}
where $\operatorname{sim}(\cdot,\cdot)$ denotes the cosine similarity score, $\mathbb{I}(k \neq i)$ is an indicator function, and $\tau$ is a temperature parameter. The overall contrastive loss for this mini-batch is the sum of $\ell_{i, j}$ over all  positive pairs. Then SimCLR minimizes the overall contrastive loss via jointly pre-training the encoder $f$ and the projection head $h$ by back-propagation.

\myparatight{MoCo~\cite{he2020momentum}} In addition to  data augmentation, MoCo has three key components: a query encoder $f_q$, a momentum encoder $f_m$, and a dictionary $\mathcal{D}$. Given an input $\mathbf{x}$, the query encoder $f_q$ and the momentum encoder $f_m$ output feature vectors $f_q(\mathbf{x})$ and $f_m(\mathbf{x})$ for it, respectively. In particular, the query encoder $f_q$ and the momentum encoder $f_m$ have the same encoder architecture. The feature vectors outputted by the momentum encoder $f_m$ are also called \emph{keys}. The dictionary $\mathcal{D}$ maintains a queue of keys via adding keys that are produced in the most recent mini-batches while removing those oldest keys from it. MoCo updates the momentum encoder $f_m$ much slower than the query encoder $f_q$ to maintain the consistency of keys in the dictionary $\mathcal{D}$. Similar to SimCLR, given a mini-batch of $K$ pre-training inputs $\{\mathbf{x}_{1}, \mathbf{x}_{2}, \cdots, \mathbf{x}_{K}\}$, the data augmentation module generates two randomly augmented views for each of them. 
For each positive pair ($\mathbf{x}_{i}^{\prime}, \mathbf{x}_{j}^{\prime}$), $\mathbf{x}_{i}^{\prime}$ and $\mathbf{x}_{j}^{\prime}$ are fed to the query encoder $f_q$ and momentum encoder $f_m$, respectively. The key $f_m(\mathbf{x}_{j}^{\prime})$ is enqueued into the dictionary $\mathcal{D}$. Given a positive pair $(\mathbf{x}_i^{\prime},\mathbf{x}_j^{\prime})$, the contrastive loss of MoCo is defined as follows:
\begin{align}
\ell_{i,j}=-\log \left(\frac{\exp \left(\operatorname{sim}\left(f_q(\mathbf{x}_i^{\prime}), f_m(\mathbf{x}_j^{\prime})\right) / \tau\right)}{\sum_{\mathbf{d}\in \mathcal{D}} \exp \left(\operatorname{sim}\left(f_q(\mathbf{x}_i^{\prime}), \mathbf{d}\right) / \tau\right)}\right),
\end{align}
where $\operatorname{sim}(\cdot,\cdot)$ means  cosine similarity  and $\tau$ is a temperature parameter. The overall contrastive loss for this mini-batch is the sum of $\ell_{i, j}$ over all positive pairs. MoCo minimizes the overall contrastive loss via pre-training the query encoder $f_q$ by back-propagation and then updates the momentum encoder $f_m$ in a slowly evolving manner. The learnt query encoder is used as a pre-trained encoder for downstream tasks.

\subsection{Training Downstream Classifiers}
The pre-trained encoder is used as a general-purpose feature extractor for various downstream tasks. In particular, given a pre-trained  encoder and a small amount of downstream training data for a downstream task, the  encoder outputs feature vectors for the training inputs. Then, a downstream classifier is trained based on the extracted feature vectors and corresponding labels through standard supervised learning. In the testing phase, the  encoder is first used to extract feature vectors for testing inputs. Then the downstream classifier outputs predicted labels for the extracted feature vectors.
\section{Problem Formulation}
\subsection{Threat Model}

\myparatight{Attacker's goal} An attacker selects one or multiple \emph{target downstream tasks}. For each target downstream task $t$, the attacker chooses $k_t$ \emph{target inputs}. We use $x_{ti}$ to denote the $i$-th target input for the $t$-th target downstream task, where $i=1,2,\cdots,k_t$ and $t=1, 2,\cdots, T$. For each target input $x_{ti}$, the attacker chooses a target class $y_{ti}$, which is different from the true class of $x_{ti}$. We note that multiple target inputs could have the same target class, i.e., the attacker desires these target inputs to be classified as the same target class. The attacker aims to inject poisoning inputs into the unlabeled pre-training dataset, such that a poisoned encoder is pre-trained and the downstream classifiers trained based on the poisoned encoder for the target downstream tasks should simultaneously classify the attacker-chosen target inputs as the corresponding attacker-chosen target classes.  

\myparatight{Attacker’s background knowledge}
We assume the attacker only has access to some images (called \emph{reference inputs}) from each target class. We denote the set of reference inputs from target class $y_{ti}$ as $X_{y_{ti}}$. We note that the reference inputs are not used to train the downstream classifiers for the target downstream tasks. The attacker can collect the reference inputs from different sources, e.g., from the Internet. \SHF{Moreover, we assume the attacker does not know the encoder's pre-training dataset, architecture, nor loss function.}

\myparatight{Attacker's capability}
The attacker is  able to inject poisoning images into the unlabeled pre-training dataset before it is used to pre-train an encoder. An encoder provider often automatically collects a large pre-training dataset from the Internet using a web crawler and uses it to pre-train an  encoder. Therefore, the attacker can publish the poisoning images on the Internet, e.g., host them on some crawler-accessible websites or post them on social medias, so an encoder provider would collect them as a part of its pre-training dataset. We consider the attacker has resources to inject  at most $N$ poisoning images/inputs into the pre-training dataset. For simplicity, we denote  the clean pre-training dataset as $X_c$ and we denote the set of poisoning inputs as $X_{p}$. 

\subsection{Data Poisoning Attack}
Given the attacker’s goal, background knowledge, and capability, we formulate our data poisoning attack as follows. Recall that the attack requires  downstream classifiers built based on the poisoned encoder to classify the target inputs as the corresponding target classes. However, it is challenging to directly quantify this goal using downstream classifiers, as we assume the attacker does not have control over the training of downstream classifiers in our threat model. To address the challenge, we propose to quantify the goal using the feature vectors outputted by the poisoned encoder. Intuitively, if the poisoned encoder produces similar feature vectors for a target input $x_{ti}$ with a target class $y_{ti}$ and reference inputs from the target class $y_{ti}$, then a downstream classifier built based on the poisoned encoder would be likely to classify the target input $x_{ti}$ as the target class $y_{ti}$. 

Therefore, the poisoned encoder should produce similar feature vectors for the target inputs and reference inputs in the same target class. In particular, we use a loss term $\mathcal{L}_{\mathrm{sim}}(x_{ti}, x_r; \theta)$ to quantify feature similarity between a target input $x_{ti}$ and a reference input $x_r$, where their feature vectors are outputted by an encoder $\theta$ and the feature similarity is the \emph{cosine similarity} between the two feature vectors. We adopt cosine similarity because it is  used by contrastive learning to measure feature similarity.  Our goal is to construct the poisoning inputs $X_p$ such that an encoder pre-trained on the poisoned pre-training dataset $X_c \cup X_p$ maximizes the total loss for all target inputs and reference inputs. Formally, we have the following optimization problem:
\begin{align}
\label{bi-opt-outer}
	\underset{X_{p}}{\operatorname{max}}\frac{1}{\sum_{t=1}^{T}\sum_{i=1}^{k_t} |X_{y_{ti}}|}  \sum_{t=1}^{T}\sum_{i=1}^{k_t} \sum_{x_r\in X_{y_{ti}}} & \mathcal{L}_{\mathrm{sim}}\left(x_{ti}, x_r; \theta^{*}\left(X_c \cup X_p\right)\right), \\
	\label{bi-opt-inner}
	s.t.\ \theta^{*}\left(X_c \cup X_p\right)=\underset{\theta}{\operatorname{argmin}} & \mathcal{L}_{\text {CL}}\left(X_{c} \cup X_{p} ; \theta\right),
\end{align}
where $T$ is the number of target downstream tasks, $k_t$ is the number of target inputs for target downstream task $t$, $X_{y_{ti}}$ is the set of reference inputs from the target class $y_{ti}$, $\theta^{*}\left(X_c \cup X_p\right)$ is the poisoned encoder pre-trained on the poisoned pre-training dataset $X_c \cup X_p$, Equation~\ref{bi-opt-outer} means that the poisoned encoder should produce similar feature vectors for the target inputs and the corresponding reference inputs, and Equation~\ref{bi-opt-inner} means that the poisoned encoder is pre-trained on the poisoned pre-training dataset using contrastive learning, where $\mathcal{L}_{\text {CL}}$ is the contrastive loss. 

The set of poisoning inputs  $X_{p}$ is a solution to the above optimization problem. We note that the optimization problem  is a \emph{bilevel} one, where Equation~\ref{bi-opt-outer} is the \emph{outer optimization} and Equation~\ref{bi-opt-inner} is the \emph{inner optimization}. 

\begin{figure*}[!t]
\vspace{-2mm}
    \centering
    {\includegraphics[width=0.9\textwidth]{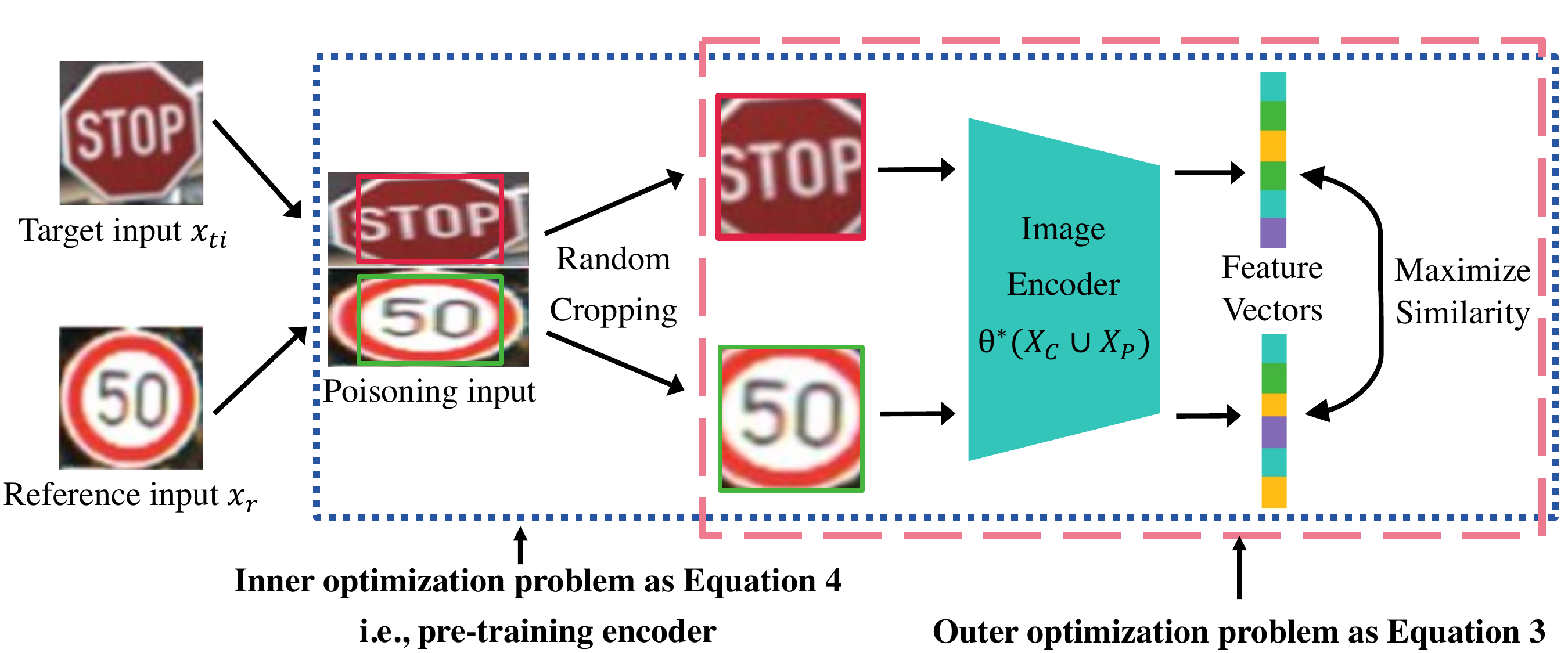}}
    \caption{Illustration of PoisonedEncoder. Solving the inner optimization problem on the poisoned pre-training data also approximately maximizes the outer objective function.}
    \label{fig-illustration}
\end{figure*}

\section{Our PoisonedEncoder}
\subsection{Challenges}

The key of our data poisoning attack is to construct a set of poisoning inputs, which is a solution to the bilevel optimization problem in Equation~\ref{bi-opt-outer}. 
However, it is notoriously challenging to solve bilevel optimization problems in general~\cite{munoz2017towards}. Specifically, gradient descent is a well-known iterative method to solve optimization problems. Given an initial set of poisoning inputs $X_p$, we calculate the gradient of Equation~\ref{bi-opt-outer} with respect to  $X_p$ (i.e., $\frac{\partial \sum_{t=1}^{T}\sum_{i=1}^{k_t} \sum_{x_r\in X_{y_{ti}}}\mathcal{L}_{\mathrm{sim}}\left(x_{ti}, x_r; \theta^{*}\left(X_c \cup X_p\right)\right)}{\sum_{t=1}^{T}\sum_{i=1}^{k_t} |X_{y_{ti}}|\cdot \partial X_p}$), and then we move $X_p$ along the gradient with a small step. However,  calculating such gradient requires the gradient of the poisoned encoder $\theta^{*}\left(X_c \cup X_p\right)$ with respect to the poisoning inputs $X_p$, i.e., $\frac{\partial {\operatorname{argmin}}_{\theta} \mathcal{L}_{\text {CL}}\left(X_{c} \cup X_{p} ; \theta\right)}{\partial X_p}$. In our threat model,  the attacker faces multiple challenges when calculating such gradient: 1) the attacker does not know the clean pre-training dataset $X_c$, 2) the attacker does not know the contrastive loss function $\mathcal{L}_{\text {CL}}$, and 3) the encoder is a highly non-linear deep neural network. As a result, it is  challenging to use iterative method to solve the bilevel optimization problem.

Therefore, we resort to non-iterative heuristic solutions to the bilevel optimization problem.  
In particular, we propose an approximate solution that is tailored to contrastive learning. 
Our approximate solution does not require the gradient of the poisoned encoder $\theta^{*}\left(X_c \cup X_p\right)$ with respect to the poisoning inputs $X_p$. Moreover, our solution does not require
access to the clean pre-training dataset nor the contrastive loss function, and is applicable to highly non-linear deep neural network based encoder. Next, we first introduce the intuition of our approximate solution, and then describe the details on constructing poisoning inputs. 
\begin{figure}[!t]
    \centering
    \subfloat[Combination 1]{
      \includegraphics[width=0.2\textwidth]{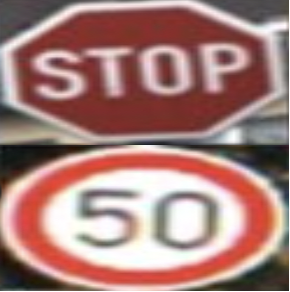}
    }
    \subfloat[Combination 2]{
      \includegraphics[width=0.2\textwidth]{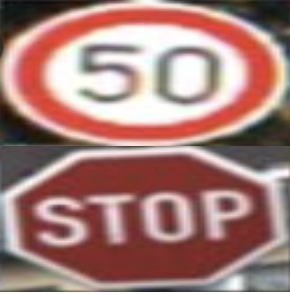}
    }
    
    \subfloat[Combination 3]{
      \includegraphics[width=0.2\textwidth]{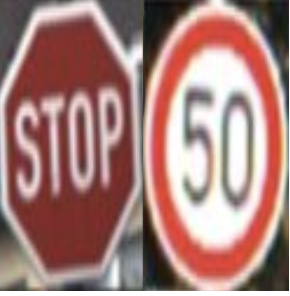}
    }
    \subfloat[Combination 4]{
      \includegraphics[width=0.2\textwidth]{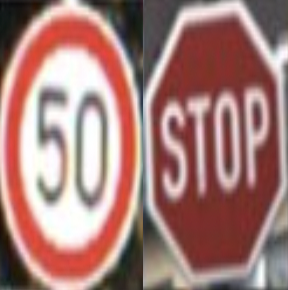}
    } 
    \caption{Four combination methods to construct poisoning inputs from a target input and a reference input.}
    \label{fig:combinations}
\end{figure}

\subsection{Our Intuition}

The key idea of our method is to construct poisoning inputs such that solving the inner optimization also approximately solves the outer optimization.  
Note that an encoder provider essentially solves the inner optimization problem when pre-training an encoder based on a poisoned pre-training dataset. Therefore, in our attack,  an encoder pre-trained on a poisoned pre-training dataset approximately maximizes the outer objective function. 
Contrastive learning aims to learn an encoder that produces similar feature vectors for two randomly cropped views of an input. In other words, the inner optimization in Equation~\ref{bi-opt-inner} aims to optimize a poisoned encoder that outputs similar feature vectors for any two randomly cropped views of an input in the poisoned pre-training dataset $X_c \cup X_p$. The outer optimization in Equation~\ref{bi-opt-outer} aims to maximize the similarity between a target input and a reference input. 
Therefore, if two randomly cropped views of a poisoning input correspond to a target input and a reference input, then the poisoned encoder obtained from the inner optimization would also maximize the similarity loss in the outer optimization. 

Based on this intuition, our PoisonedEncoder constructs a poisoning input by combining a target input and a reference input, which is illustrated in Figure~\ref{fig-illustration}. Moreover, to increase diversity of the poisoning inputs, we consider four ways of combining a target input and a reference input, i.e.,  up (target)-down(reference), up (reference)-down(target), left (target)-right(reference), and left (reference)-right(target),  which are illustrated in Figure~\ref{fig:combinations}. As our experiments will show, PoisonedEncoder achieves higher attack success rate by using all four combination methods. We note that such combined images also naturally appear on the Internet. Figure~\ref{figures-combined-google-search} in Appendix shows some examples of combined images from Google search.

\subsection{Constructing Poisoning Inputs}

\begin{algorithm}[!t]
    \caption{Our PoisonedEncoder}
    \begin{algorithmic}[1]
    \STATE {\bfseries Input:} Target downstream tasks $t=1,2, \cdots, T$, target inputs $x_{ti}$ and target classes $y_{ti}$ where $i=1,2, \cdots, k_{t}$, reference inputs $X_{y_{ti}}$ from the target class $y_{ti}$, number of poisoning inputs $N$, a set of combination methods $\mathcal{M}$.\\
    \STATE {\bfseries Output:} Poisoning inputs $X_P$ \\
    \STATE $X_P= \emptyset $ $\backslash\backslash$ Initialize $X_P$ as an empty set \\
    \WHILE {|$X_P$| $\leq N$}
    \STATE Randomly pick a target input $x_{ti}$\\
    \STATE Randomly pick a reference input $x_{r} \in X_{y_{ti}}$ \\
    \STATE Randomly pick a combination method $m\in \mathcal{M}$ \\
    \STATE Combine $x_{ti}$ and $x_{r}$ based on $m$ as poisoning input \\
    \STATE Add the poisoning input to $X_P$\\
    \ENDWHILE
    \STATE \textbf{return} $X_P$\\
    \end{algorithmic}
\label{algorithml1}
\end{algorithm}

Algorithm~\ref{algorithml1} shows the algorithmic details of \SHF{PoisonedEncoder}. Given a set of target downstream tasks, target inputs and target classes for each target downstream task, reference inputs from each target class, and a set of combination methods. 
We construct $N$ poisoning inputs. To construct a poisoning input, we randomly pick a target input with a target class, randomly pick a reference input from the target class, and randomly pick a combination method; and then we combine the target input and the reference input according to the combination method as a poisoning input.   

\section{Evaluation}

\subsection{Experimental Setup}
\label{exp-setup}

\myparatight{Pre-training datasets and encoders} 
We consider two benchmark datasets, CIFAR10~\cite{krizhevsky2009learning} and Tiny-ImageNet~\cite{tinyimagenet}, as pre-training datasets. CIFAR10 dataset contains 50,000 training images and 10,000 testing images. Tiny-ImageNet dataset contains 100,000 training images and 10,000 testing images. Following prior work~\cite{chen2020simple,he2020momentum}, we use the training images of each dataset as a pre-training dataset \SHF{and we use multiple data augmentations (random cropping, color jittering, Gaussian blurring, random grayscale, and random horizontal flipping) when pre-training encoders in experiments. Following prior work~\cite{chen2020simple,he2020momentum}, we consider different scales of random cropping, i.e., scale is randomly sampled from [0.08, 1] each time.} Moreover, we rescale each image to 32x32 in both datasets.

Unless otherwise mentioned, we use ResNet18~\cite{he2016deep} as the neural network architecture for an encoder and SimCLR~\cite{chen2020simple}  to pre-train both poisoned and clean encoders. In particular, we adopt the publicly available implementation~\cite{simclr-code} of SimCLR with the default settings. We pre-train an encoder for 300 epochs using the Adam optimizer, an initial learning rate of 0.001, and a batch size of 512. However, we will also explore the impact of encoder architecture, contrastive learning algorithm, learning rate, batch size, and number of pre-training epochs on PoisonedEncoder.

\myparatight{Training downstream classifiers} 
Given a pre-trained (poisoned or clean) encoder, we train downstream classifiers on three downstream datasets, STL10~\cite{coates2011analysis}, Facemask~\cite{facemask}, and EuroSAT\SHF{~\cite{helber2019eurosat}}. STL10 contains 13,000 labeled color images from 10 classes. Specifically, the labeled data is divided into 5,000 training images and 8,000 testing images. Facemask dataset is a binary classification task for detecting whether a person wears a mask given his/her face image. It contains 10,000 training images and 993 testing images. EuroSAT dataset consists of 10 classes with 27,000 labeled geographical images collected by satellites. We randomly and evenly split the images into training and testing sets in EuroSAT. Like the pre-training data, we also rescale each image to 32x32 in these downstream datasets. 

Following the contrastive learning paradigm~\cite{chen2020simple,he2020momentum}, we adopt a neural network with one fully-connected layer as a downstream classifier. Moreover, we train a downstream classifier for 100 epochs using Adam optimizer, an initial learning rate of 0.001, and a batch size of 512. We will explore the impact of learning rate, batch size, and number of epochs used to train a downstream classifier on PoisonedEncoder.

\myparatight{Evaluation metrics}
We use  \emph{attack success rate (ASR)} as well as a downstream classifier's \emph{clean accuracy (CA)} and \emph{poisoned accuracy (PA)} to evaluate PoisonedEncoder. ASR is used to measure the attack success of PoisonedEncoder, while CA and PA are used to measure the impact of \SHF{PoisonedEncoder} on the utility of the encoder. ASR is the fraction of target inputs that are predicted as the corresponding target classes by the target downstream classifiers trained on a poisoned encoder. A higher ASR indicates a more successful  attack. A downstream classifier's CA and PA are the testing accuracy of the downstream classifier trained on a clean and poisoned encoder, respectively. A smaller difference between CA and PA indicates that the attack better preserves the utility of the encoder.

\myparatight{Compared data poisoning attacks} We compare our PoisonedEncoder with the following two baseline attacks. Note that these attacks were originally designed for supervised learning and semi-supervised learning, and we extend them to contrastive learning. 

\begin{itemize}

\item \myparatight{Witches' Brew~\cite{geiping2020witches}} Witches' Brew is a state-of-the-art targeted data poisoning attack to deep neural network classifier. The  attack is formulated as a bilevel optimization problem. 
Roughly speaking, Witches' Brew uses a gradient-based iterative method to solve the bilevel optimization problem. In each iteration, they update the poisoning inputs to maximize the alignment between the gradient of the inner objective function and that of the outer objective function, where the gradient is with respect to the parameters of a clean classifier. 
We extend their method to solve our formulated bilevel optimization problem to craft poisoning inputs. Specifically, we optimize poisoning inputs to maximize the alignment between $ \sum_{t=1}^{T} \sum_{i=1}^{k_{t}} \sum_{x_{r} \in X_{y_{t i}}} -\nabla_{\theta} \mathcal{L}_{\operatorname{sim}}\left(x_{t i}, x_{r} ; \theta\left(X_{c} \cup X_{p}\right)\right)$ and  $  \nabla_{\theta}\mathcal{L}_{\mathrm{CL}}\left(X_{c} \cup X_{p} ; \theta\right)$, where $\theta$ is a clean encoder. 
Moreover, following Withes' Brew, we replace $\nabla_{\theta} \mathcal{L}_{\mathrm{CL}}\left(X_{c} \cup X_{p} ; \theta\right)$ as $\nabla_{\theta} \mathcal{L}_{\mathrm{CL}}\left(X_{p} ; \theta\right)$.   Witches' Brew  requires the attacker has access to a clean encoder and the contrastive loss function $\mathcal{L}_{\mathrm{CL}}$, where we assume the clean encoder is trained on a clean pre-training dataset in our experiments. Note that we give  advantages to Witches' Brew, as our PoisonedEncoder does not require these information.

\item \myparatight{\SHF{Interpolation Consistency Poisoning (ICP)}~\cite{carlini2021poisoning}} Carlini proposed \SHF{ICP}, a targeted data poisoning attack, to semi-supervised learning. 
Roughly speaking, \SHF{ICP}  crafts unlabeled poisoning inputs as interpolations between a target input and reference inputs. We apply \SHF{ICP} to contrastive learning.
For a fair comparison, \SHF{PoisonedEncoder} and \SHF{ICP} use the same target input and reference inputs to craft poisoning inputs. We implemented \SHF{ICP} and adopted its default parameter settings from the paper. 
    
\end{itemize}

\begin{table}[!t]\renewcommand{\arraystretch}{1.2} 
	\fontsize{8}{8}\selectfont
	\centering
	\caption{ASRs of different attacks.}
	\setlength{\tabcolsep}{1mm}
	{
	\begin{tabular}{|c|c|c|c|c|}
		\hline
	\makecell{Pre-training \\Dataset} & \makecell{Target Downstream \\Dataset} & \makecell{Witches' Brew} & \SHF{\makecell{ICP}} & \makecell{Ours} \\ \hline
	\multirow{3}{*}{CIFAR10}
	& STL10     & 0.1   &  0.5 &  0.8\\ \cline{2-5}  
	& Facemask  & 0.1   & 0.6 & 0.9 \\ \cline{2-5}  
	& EuroSAT  & 0.0  & 0.2 & 0.5\\ \hline 
	\multirow{3}{*}{Tiny-ImageNet} 
	& STL10   &  0.0   & 0.4   & 0.7 \\ \cline{2-5}  
	& Facemask  & 0.1  &  0.8 & 1.0 \\ 
	\cline{2-5}  
	& EuroSAT & 0.0   &  0.2 &0.4\\ \hline  
	\end{tabular}
	}
	\label{table_asr}
	\vspace{-2mm}
\end{table}

\myparatight{Parameter settings} Unless otherwise mentioned, we consider the following parameter settings for PoisonedEncoder: the attacker chooses one target downstream dataset, randomly picks one testing input from the target downstream dataset as a target input, and randomly picks a class that is not the true class of the target input as a target class. The attacker has 50 reference inputs that are testing inputs randomly sampled from the target class in the target downstream dataset. The attacker injects 1\% poisoning inputs to a pre-training dataset. In particular, the attacker injects 500 and 1,000 poisoning inputs into the pre-training datasets CIFAR10 and Tiny-ImageNet, respectively. 
For each experiment, we repeat it for 10 trials and report the average results. Unless otherwise mentioned, we assume the target downstream dataset is STL10. We note that after combining a target input and a reference input as a poisoning input, we resize the poisoning input to have the same size as the target/reference input, i.e., 32x32.  \SHF{We performed experiments on 18 NVIDIA-RTX-6000 GPUs, each of which has 24 GB memory.}

\subsection{Experimental Results}
\myparatight{PoisonedEncoder achieves high attack success rates} Table~\ref{table_asr} shows the ASRs of different attacks on different pre-training datasets and target downstream datasets.  Our PoisonedEncoder achieves higher ASRs than the compared attacks. Specifically, our attack achieves at least 0.2 higher ASR than \SHF{ICP}, while Witches' Brew is almost ineffective. The reason is that \SHF{ICP} and Witches' Brew were respectively tailored for semi-supervised learning and supervised learning,  and they achieve suboptimal success rates when extended to contrastive learning. 

In particular, the poisoning inputs constructed by \SHF{PoisonedEncoder} are better solutions to our formulated bilevel optimization problem in Equation~\ref{bi-opt-outer} and~\ref{bi-opt-inner} than those constructed by  Witches' Brew and \SHF{ICP}. To further illustrate this point, we calculate the cosine similarity score between the feature vector of the target input and that of each reference input, and we compute the average of the cosine similarity scores, which is the value of the outer objective function in Equation~\ref{bi-opt-outer}.    Table~\ref{table-similarity} shows the average cosine similarity scores (i.e., the outer objective function values) under different attacks. We observe that PoisonedEncoder achieves higher average cosine similarity score, which confirms that PoisonedEncoder is a better solution to the bilevel optimization problem than Witches' Brew and \SHF{ICP}.

We also observe that \SHF{ICP} achieves moderate ASRs and is much more successful than Witches' Brew. This is because both the semi-supervised learning algorithms attacked by \SHF{ICP} and contrastive learning use data augmentations during (pre-)training. As a result,  \SHF{ICP} is a better solution to our formulated bilevel optimization problem than Witches' Brew, as shown in Table~\ref{table-similarity}. 

\SHF{Our PoisonedEncoder is also effective when the pre-training dataset is large. For instance, our attack's ASR is 0.7 when the pre-training dataset is the full ImageNet dataset~\cite{russakovsky2015imagenet} and the target downstream dataset is Facemask, where the parameters are set to their default settings. It took us around 5 days to pre-train 10 encoders on the full ImageNet dataset. Due to the limited computing resources, we were not able to perform other experiments on the full ImageNet dataset. }

\begin{table}[!t]\renewcommand{\arraystretch}{1.3} 
	\fontsize{8}{8}\selectfont
	\centering
	\caption{Values of the outer objective function in Equation~\ref{bi-opt-outer} obtained under different attacks, where pre-training dataset is CIFAR10 and target downstream dataset is STL10.}
	\setlength{\tabcolsep}{1mm}
	{
	\begin{tabular}{|c|c|c|c|}
		\hline
	\makecell{No Attack} & \makecell{Witches' Brew} & \SHF{\makecell{ICP}} & \makecell{Ours} \\ \hline
	0.183 & 0.257  &  0.463   & 0.689 \\ \hline  
	\end{tabular}
	}
	\label{table-similarity}
\end{table}

\begin{table}[!t]\renewcommand{\arraystretch}{1.2} 
	\fontsize{7.5}{8}\selectfont
	\centering
	\caption{CAs and PAs of downstream classifiers. }
	\setlength{\tabcolsep}{1mm}
	{
	\begin{tabular}{|c|c|c|c|c|}
		\hline
	\makecell{Pre-training \\Dataset} & \makecell{Target Downstream \\Dataset} & \makecell{Downstream \\Dataset}  &  \makecell{CA } & \makecell{PA} \\ \hline
	\multirow{9}{*}{CIFAR10} & \multirow{3}{*}{STL10} & STL10     &  0.718   &  0.715\\
	\cline{3-5}  
	 & & Facemask     &  0.947   &  0.952\\
	\cline{3-5}  
	& & EuroSAT     &  0.815   &  0.821\\
	\cline{2-5}  
	& \multirow{3}{*}{Facemask} & STL10     &  0.718    &  0.716\\
	\cline{3-5}  
	& & Facemask     &   0.947    & 0.937\\
	\cline{3-5}  
	& & EuroSAT     &  0.815   &  0.820\\
	\cline{2-5}  
	& \multirow{3}{*}{EuroSAT} & STL10     &  0.718    &  0.724\\
	\cline{3-5}  
	& & Facemask     &  0.947   &  0.953\\
	\cline{3-5}  
	& & EuroSAT    &  0.815 & 0.797 \\
	\hline
	\multirow{9}{*}{Tiny-ImageNet} & \multirow{3}{*}{STL10} & STL10    & 0.635   & 0.637  \\
	\cline{3-5}  
	 & & Facemask     &  0.965   &  0.968\\
	\cline{3-5}  
	& & EuroSAT     &  0.816   &  0.853\\
	\cline{2-5}  
	& \multirow{3}{*}{Facemask} & STL10     &   0.635   &  0.633\\
	\cline{3-5}  
	& & Facemask     & 0.965  &  0.977 \\ 
	\cline{3-5}  
	& & EuroSAT     &  0.816   &  0.855\\
	\cline{2-5}  
	& \multirow{3}{*}{EuroSAT} & STL10     &   0.635   &  0.633\\
	\cline{3-5}  
	& & Facemask     &  0.965   &  0.970\\
	\cline{3-5}  
	& & EuroSAT   & 0.816   &  0.844 \\
	\hline
	\end{tabular}
	}
	\label{table_accuracy}
\end{table}

\begin{table}[!t]\renewcommand{\arraystretch}{1.3} 
	\fontsize{8}{8}\selectfont
	\centering
    \caption{\SHF{ASR of PoisonedEncoder when using different combination methods to construct poisoning inputs. The combination methods come from Figure~\ref{fig:combinations}.}}
	\subfloat[\SHF{Pre-trained on CIFAR10} ]
	{\setlength{\tabcolsep}{1mm}
	{
	\begin{tabular}{|c|c|}
		\hline
	\SHF{\makecell{Combination Method}} & \SHF{\makecell{ASR}}  \\ \hline
	\SHF{1}
	& \SHF{0.4}     \\ \hline
    \SHF{2}   &  \SHF{0.5}\\ \hline
    \SHF{3}   &  \SHF{0.5}\\ \hline
    \SHF{4}   &  \SHF{0.6}\\ \hline
    \SHF{1+2}   &  \SHF{0.5}\\ \hline
    \SHF{1+2+3}   &  \SHF{0.6}\\ \hline
    \SHF{1+2+3+4}   &  \SHF{0.8}\\ \hline
	\end{tabular}
	}
	\label{table_diff_combination_cifar}}
	\subfloat[\SHF{Pre-trained on Tiny-ImageNet}]
	{\setlength{\tabcolsep}{1mm}
	{
	\begin{tabular}{|c|c|}
		\hline
	\makecell{\SHF{Combination Method}} & \makecell{\SHF{ASR}} \\ \hline
	\SHF{1}
	& \SHF{0.3}     \\ \hline
    \SHF{2}   &  \SHF{0.3}\\ \hline
    \SHF{3}   &  \SHF{0.2}\\ \hline
    \SHF{4}   &  \SHF{0.4}\\ \hline
    \SHF{1+2}   &  \SHF{0.5}\\ \hline
    \SHF{1+2+3}   &  \SHF{0.6}\\ \hline
    \SHF{1+2+3+4}   &  \SHF{0.7}\\ \hline
	\end{tabular}
	}
	\label{table_diff_combination_pre_tiny}}
        \label{table_diff_combination}
\end{table}

\begin{figure}[!t]
\vspace{-7mm}
    \centering
    \subfloat[]{\label{figure-impact-ref-imgs}
      \includegraphics[width=0.23\textwidth]{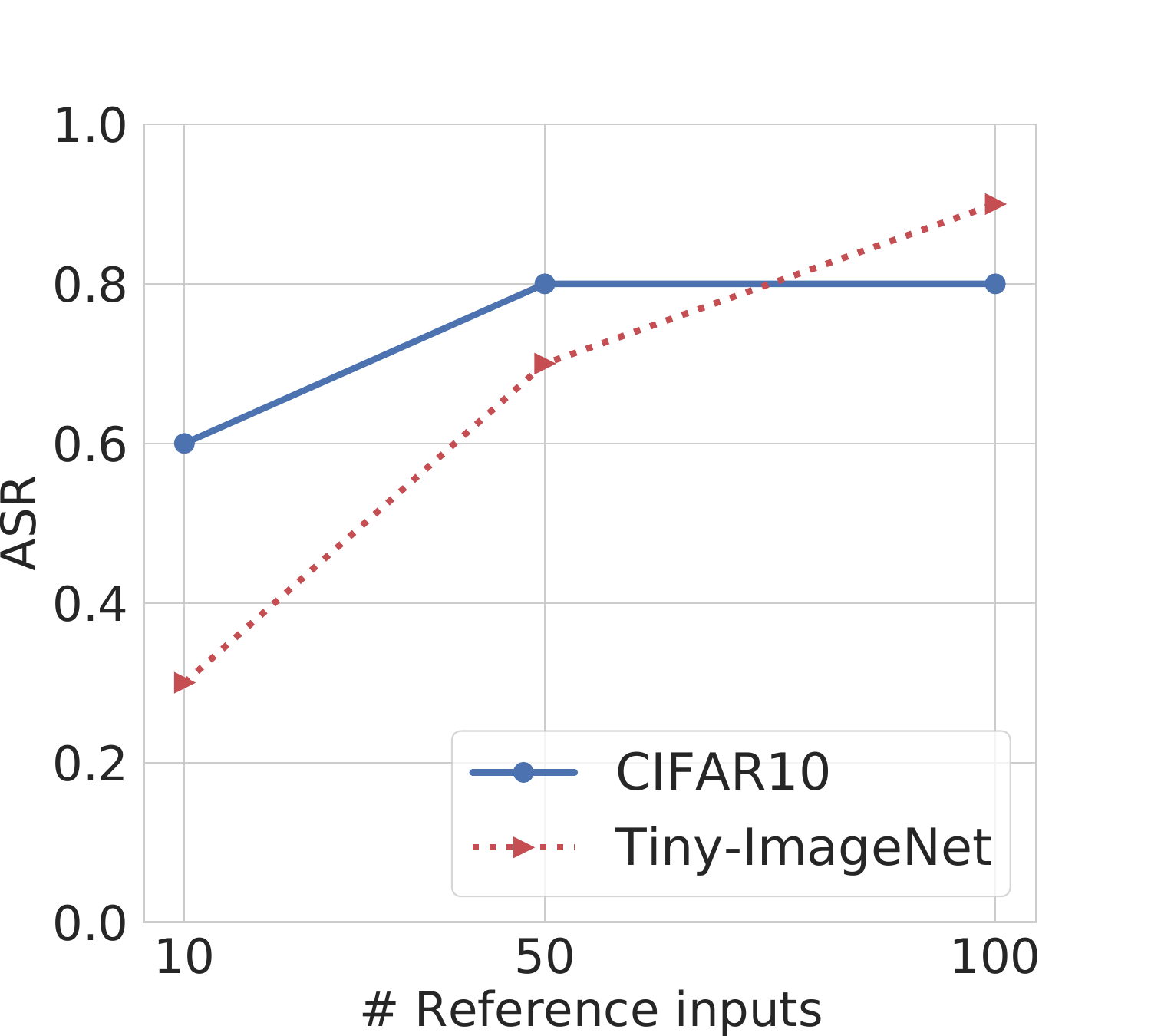}
    }
    \subfloat[]{\label{fig-impact-of-poisoning-rate}
    \includegraphics[width=0.23\textwidth]{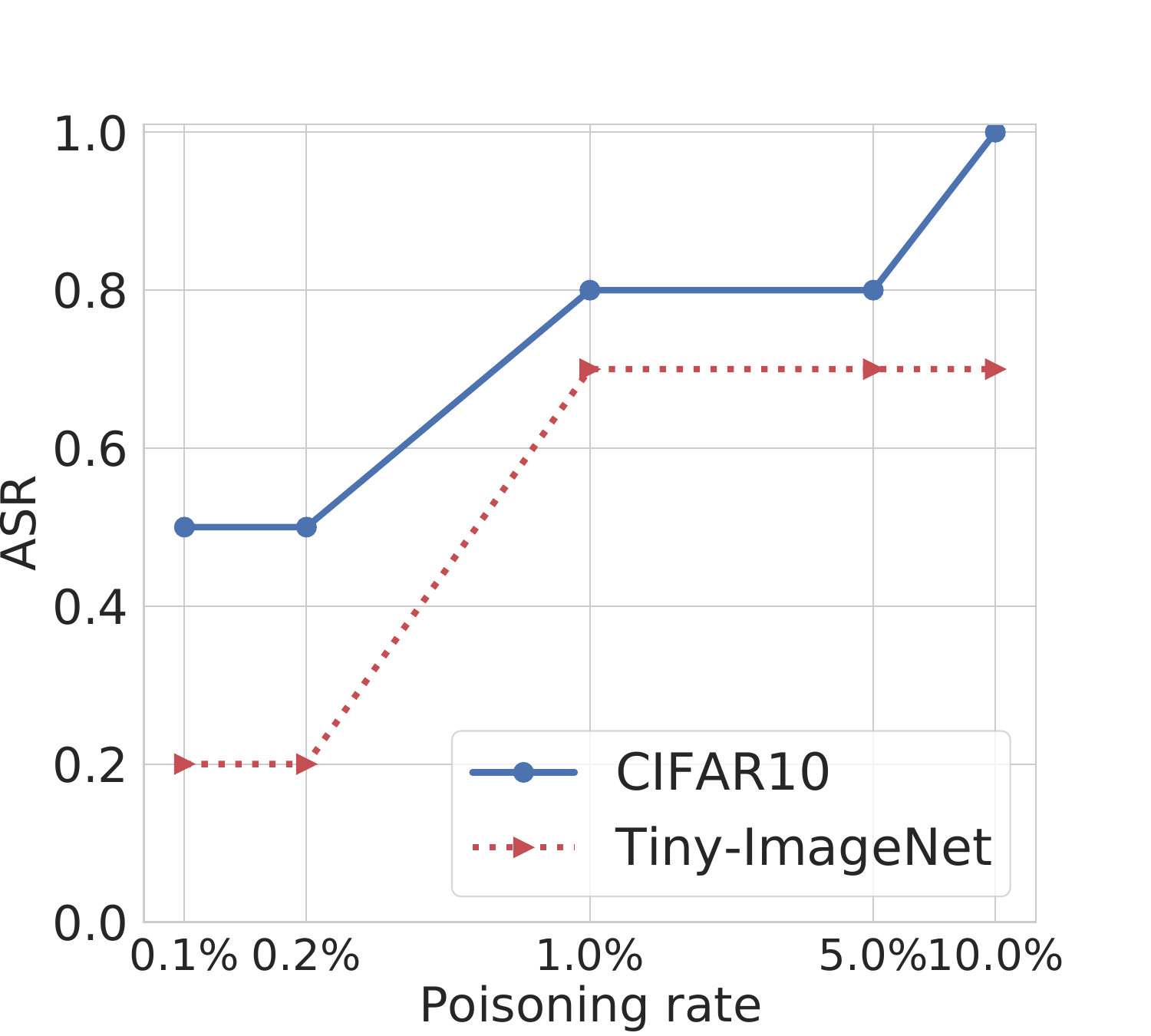}
    }
    \caption{(a) Impact of the number of reference inputs. (b) Impact of poisoning rate, where the x-axis is in log-scale.}
\end{figure}

\myparatight{PoisonedEncoder preserves utility} Table~\ref{table_accuracy} shows the CA and PA of different downstream classifiers. In particular, given a pre-training dataset and a target downstream dataset, we evaluate both CA and PA on all the three downstream datasets. 
In most cases, the gap between the CA and PA for a downstream dataset is within 1\%. Our results indicate that PoisonedEncoder preserves the utility of the encoder. This is because our poisoning inputs are still natural images and a neural network encoder is expressive, which not only learns high-performing feature representations for non-target inputs  but also learns the hidden behavior from our attack. \SHF{We also found PoisonedEncoder preserves the utility for the target class and class of the target input. For instance, when the pre-training dataset is CIFAR10 and the target downstream dataset is STL10, the CA and PA of the downstream classifier for the target class (or class of the target input) are respectively 0.702 and 0.698 (or 0.742 and 0.745).}

\myparatight{Impact of different combination methods} \SHF{PoisonedEncoder} uses four combination methods to construct poisoning inputs. One natural question is whether it is necessary to use all of them.  \SHF{Table~\ref{table_diff_combination}} shows the \SHF{ASR of PoisonedEncoder}
when \SHF{using} any one of the four combination methods or their combinations, e.g., 1+2+3 means using the first three combination methods. Our results show that PoisonedEncoder with any of the four combination methods achieves similar ASRs. When  using more combination methods,  PoisonedEncoder achieves higher ASRs, e.g.,  1+2+3 achieves higher ASRs than  1+2, and  1+2+3+4 achieves higher ASRs than  1+2+3. This is because the poisoning inputs are more diverse when more combination methods are used. 

\myparatight{Impact of the number of reference inputs and poisoning rate} Figure~\ref{figure-impact-ref-imgs} and  Figure~\ref{fig-impact-of-poisoning-rate} respectively show the impact of the number of reference inputs and poisoning rate on ASR of PoisonedEncoder for the two pre-training datasets.  A general trend is that the ASR of our PoisonedEncoder first increases and then saturates as the attacker uses more reference inputs or injects more poisoning inputs (i.e., poisoning rate increases). This is because the pre-trained poisoned encoder is more likely to produce similar feature vectors for the target inputs and some reference inputs when the number of reference inputs or the poisoning rate is larger.

\begin{table}[!t]\renewcommand{\arraystretch}{1.3} 
	\fontsize{8}{8}\selectfont
	\centering
    \caption{ASR of PoisonedEncoder for different contrastive learning algorithms.}
	\subfloat[Pre-trained on CIFAR10 ]
	{\setlength{\tabcolsep}{1mm}
	{
	\begin{tabular}{|c|c|}
		\hline
	\makecell{Pre-training Algorithm} & \makecell{ASR}  \\ \hline
	SimCLR
	& 0.8     \\ \hline
    MoCo   &  0.6\\ \hline
	\end{tabular}
	}
	\label{table_diff_pretrain_pre_cifar}}
	\subfloat[Pre-trained on Tiny-ImageNet]
	{\setlength{\tabcolsep}{1mm}
	{
	\begin{tabular}{|c|c|}
		\hline
	\makecell{Pre-training Algorithm} & \makecell{ASR} \\ \hline
	SimCLR
	& 0.7     \\ \hline
    MoCo   &  0.8\\ \hline
	\end{tabular}
	}
	\label{table_diff_pretrain_pre_tiny}}
        \label{table_diff_pretrain}
\end{table}

\begin{table}[!t]\renewcommand{\arraystretch}{1.3} 
	\fontsize{8}{8}\selectfont
	\centering
    \caption{ASR of PoisonedEncoder for different architectures.}
	\subfloat[Pre-trained on CIFAR10 ]
	{\setlength{\tabcolsep}{1mm}
	{
	\begin{tabular}{|c|c|}
		\hline
	\makecell{Encoder Architecture} & \makecell{ASR}  \\ \hline
	ResNet18
	& 0.8   \\ \hline
    VGG11   &  0.9 \\ \hline
    MobileNet-v2   &  0.8 \\ \hline
	\end{tabular}
	}
	\label{table_diff_arch_pre_cifar}}
	\subfloat[Pre-trained on Tiny-ImageNet]
	{\setlength{\tabcolsep}{1mm}
	{
	\begin{tabular}{|c|c|}
		\hline
	\makecell{Encoder Architecture} & \makecell{ASR} \\ \hline
	ResNet18
	& 0.7     \\ \hline
    VGG11   &  0.8 \\ \hline
    MobileNet-v2   &  0.8 \\ \hline
	\end{tabular}
	}
	\label{table_diff_arch_pre_tiny}}
        \label{table_diff_architecture}
\end{table}

\myparatight{PoisonedEncoder is agnostic to  contrastive learning algorithm and encoder architecture} 
Table~\ref{table_diff_pretrain} shows the ASR of PoisonedEncoder when SimCLR and MoCo are used to pre-train encoders.  For MoCo, we adopt its publicly available implementation~\cite{moco-code} with the default settings. Our results show that PoisonedEncoder is effective for both contrastive learning algorithms. This is because both algorithms use random cropping to generate augmented views during pre-training. 
Table~\ref{table_diff_architecture} shows ASR of PoisonedEncoder for different encoder architectures.  Our results show that PoisonedEncoder is agnostic to encoder architecture. This is because PoisonedEncoder does not rely on encoder architecture to construct poisoning inputs.

\begin{figure}[!t]
    \centering
    \subfloat[ASR for different true classes of target inputs]
    {\label{fig-impact-original-class}
      \includegraphics[width=0.45\textwidth]{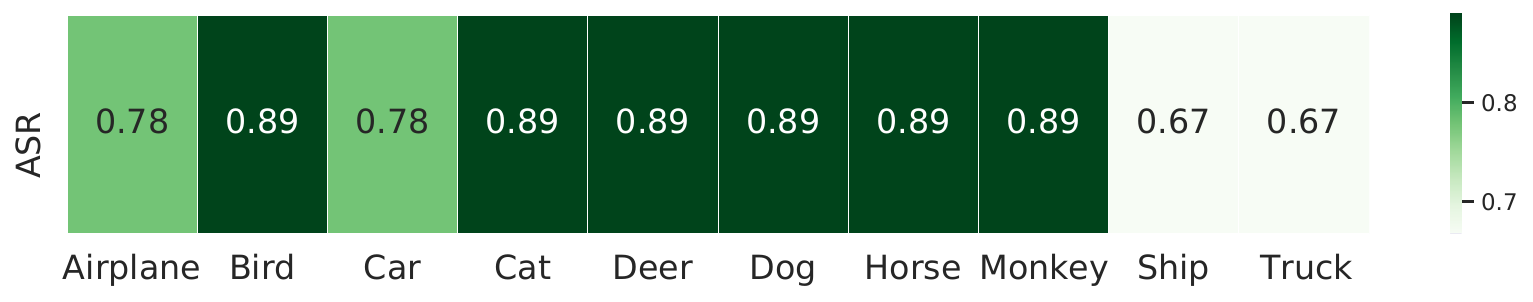}
    }
    \hfill
    \subfloat[ASR for different target classes of target inputs]{\label{fig-impact-target-class}
    \includegraphics[width=0.45\textwidth]{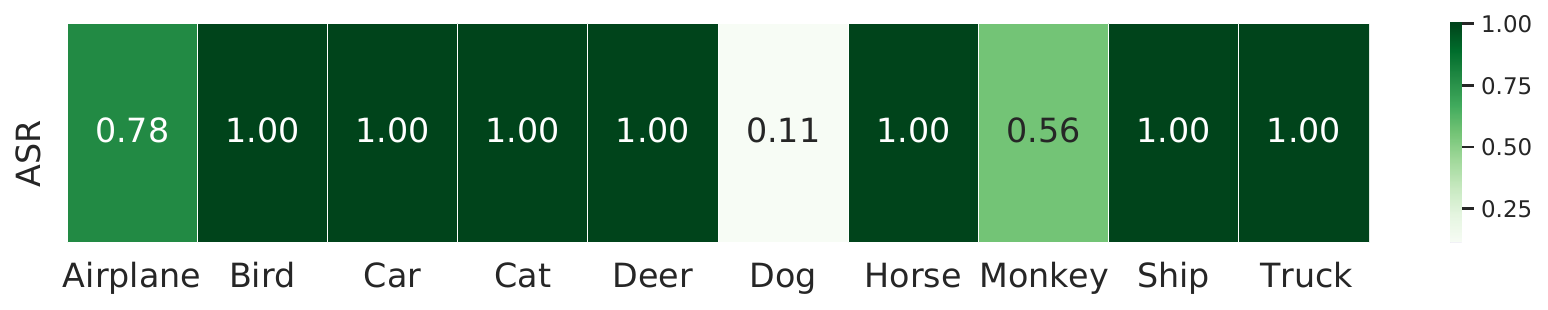}
    }
    \caption{ASR for different true or target classes.}
        \label{figure-classes}
\end{figure}

\myparatight{Impact of the true class and target class of the target input}  Figure~\ref{fig-impact-original-class} shows the ASR of PoisonedEncoder when the target input is from different true classes, where the pre-training dataset is CIFAR10 and the target downstream dataset is STL10 that has 10 classes. In particular, 
given a true class of the target downstream dataset, we have 9 target classes that are not the true class; and for each target class,  we randomly sample a target input from the true class and assign the target class for it. Therefore, we perform 9 experimental trials for a true class, and the ASR of PoisonedEncoder for the true class is averaged over the 9 trials.   Our results indicate that some true classes (e.g.,  deer, bird, dog, etc.) are easier to be attacked than other true classes (e.g.,  ship, airplane, etc.). We suspect the reason is that the pre-trained encoder's learned feature representations are more robust for some true classes, which are harder to be attacked.

Figure~\ref{fig-impact-target-class} shows the ASR of PoisonedEncoder for different target classes.  In particular, given a target class, we have 9 true classes that are not the target class; and for each true class, we randomly sample a target input from the true class and assign the target class to it. Thus, we perform 9 experimental trials for each target class, and the ASR of PoisonedEncoder for the target class is averaged over the 9 trials. 
We have two observations. First, our PoisonedEncoder achieves very high ASRs (1.00) for most target classes. This means that, for most target classes, our poisoning inputs make the poisoned encoder  output highly similar feature vectors for a target input and reference inputs, and thus the target downstream classifier trained on the poisoned encoder is very likely to predict the target input as the target class.  
\SHF{Second, we observe that the ASRs of target class `dog' (0.11) and target class `monkey' (0.56) are much lower than those of the other target classes.  We found the reason is that the target downstream classifier is less accurate for these two classes. Specifically, 
the target downstream classifier's PAs for the `dog' and `monkey' classes are respectively 0.495 and 0.542, which are much lower than the average 0.715 PA of the other classes.}

\begin{figure}[!t]
    \centering
    {\includegraphics[width=0.35\textwidth]{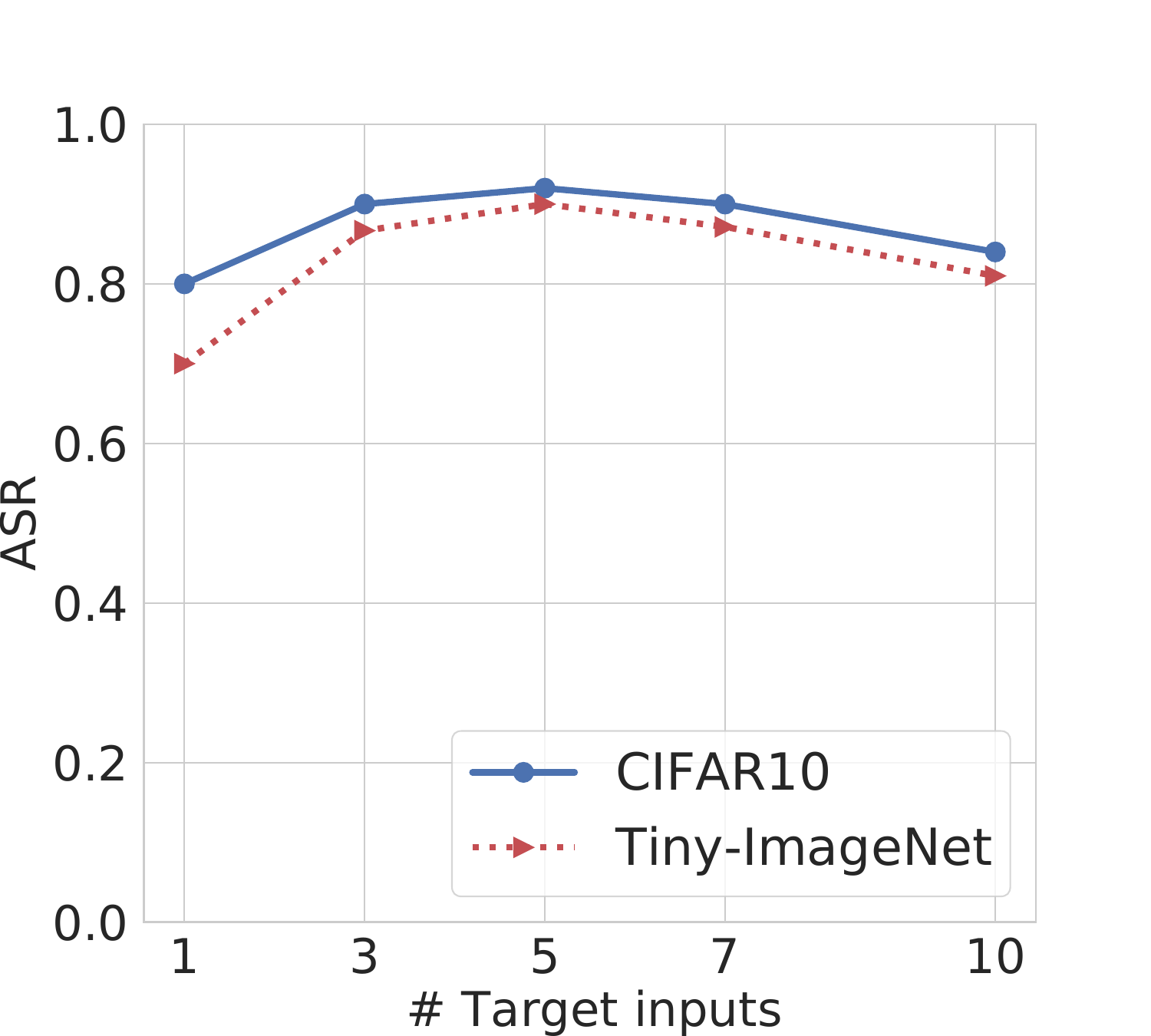}}
    \caption{ASR of PoisonedEncoder when attacking multiple target inputs simultaneously.}
    \label{fig-impact-target-inputs}
\end{figure}

\myparatight{Impact of the number of target inputs and target downstream tasks} Figure~\ref{fig-impact-target-inputs} shows the ASR of our \SHF{PoisonedEncoder} when the attacker selects different numbers of random target inputs from the same target downstream task. The ASRs keep high as the number of target inputs increases.  
Table~\ref{table_target_all} shows the ASR on each target downstream dataset when the attacker attacks three target downstream datasets simultaneously. In this experiment, for each target downstream dataset, the attacker chooses one target input and target class at random, and the poisoning rate is 1\%. We observe that PoisonedEncoder can effectively attack multiple target downstream tasks simultaneously. Moreover, by comparing Table~\ref{table_target_all} with Table~\ref{table_asr}, we find that attacking multiple target downstream tasks simultaneously achieves the same or comparable ASRs as attacking each target downstream task separately. This is because  the poisoning rate for each target downstream task is the same in these two experiments. We also found that 
attacking multiple target downstream tasks simultaneously achieves smaller ASRs when the total poisoning rate is fixed to be 1\%. In other words, ASR of PoisonedEncoder depends on the poisoning rate per target downstream task.

\begin{table}[!t]\renewcommand{\arraystretch}{1.3} 
	\fontsize{9}{8}\selectfont
	\centering
    \caption{ASR of PoisonedEncoder when attacking three target downstream tasks simultaneously.}
	\subfloat[Pre-trained on CIFAR10]
	{\setlength{\tabcolsep}{1mm}
	{
	\begin{tabular}{|c|c|c|}
		\hline
	 \makecell{Target Downstream \\Tasks} & \makecell{ASR}   \\ \hline
	STL10     &  0.8   \\ \hline  
	Facemask  &   0.9    \\ \hline    
	EuroSAT  & 0.5 \\ \hline 
	\end{tabular}
	}
	\label{table_target_all_pre_cifar}}
	\subfloat[Pre-trained on Tiny-ImageNet]
	{\setlength{\tabcolsep}{1mm}
	{
	\begin{tabular}{|c|c|}
		\hline
	 \makecell{Target Downstream \\Tasks} & \makecell{ASR}  \\ \hline
	STL10     &  0.6   \\ \hline  
	Facemask  &   1.0     \\ \hline    
	EuroSAT  & 0.4  \\ \hline 
	\end{tabular}
	}
	\label{table_target_all_pre_tiny}}
        \label{table_target_all}
\end{table}

\begin{table}[!ht]\renewcommand{\arraystretch}{1.2} 
	\centering
	\caption{ASR of PoisonedEncoder when pre-training encoder and training the target downstream classifier use different learning rates or batch sizes. }
	\setlength{\tabcolsep}{1mm}
	{
	\begin{tabular}{|c|c|c|c|}
		\hline
	\makecell{Phase} & \makecell{Parameter} & \makecell{Value}  &  \makecell{ASR }  \\ \hline
	\multirow{6}{*}{\makecell{Pre-training \\Encoders}} & \multirow{3}{*}{\makecell{Learning \\Rate}} & $5 \times 10^{-3}$   &  0.5  \\
	\cline{3-4}  
	 & & $1 \times 10^{-3}$     &  0.8   \\
	\cline{3-4}  
	& & $5 \times 10^{-4}$     &  0.8\\
	\cline{2-4}  
	& \multirow{3}{*}{Batch size} & 256    &  0.8\\
	\cline{3-4}  
	& & 512     &   0.8\\
	\cline{3-4}  
	& & 1024     &  0.8\\
	\cline{2-4}  
	\hline
	\multirow{6}{*}{\makecell{Training \\Downstream \\Classifier}} & \multirow{3}{*}{\makecell{Learning \\Rate}} & $5 \times 10^{-3}$   &  0.8\\
	\cline{3-4}  
	 & & $1 \times 10^{-3}$   &   0.8\\
	\cline{3-4}  
	& & $5 \times 10^{-4}$     &  0.8\\
	\cline{2-4}  
	& \multirow{3}{*}{Batch size} & 256    &  0.7\\
	\cline{3-4}  
	& & 512    &   0.8\\
	\cline{3-4}  
	& & 1024     &  0.8\\
	\cline{2-4}  
	\hline
	\end{tabular}
	}
	\label{parametersettings}
\end{table}

\myparatight{Impact of other parameters} Table~\ref{parametersettings} shows the ASR of PoisonedEncoder when pre-training encoder and training the target downstream classifier use different learning rates or batch sizes, where the pre-training dataset is CFIAR10 and target downstream dataset is STL10. We show the results of different number of epochs used to pre-train encoder and train downstream classifier in Section~\ref{inprocessingdefense} when we explore early stopping as a defense. PoisonedEncoder achieves consistently high ASRs across different parameter settings. PoisonedEncoder achieves a slightly lower ASR when the learning rate for pre-training encoder is $5 \times 10^{-3}$. This is because the encoder pre-trained with this learning rate outputs less-performing feature representations. In particular, the target downstream classifier has smaller CA/PA in this setting, making PoisonedEncoder achieve lower ASR. 
\section{Defenses}

\begin{figure}[!t]
    \centering
    {\includegraphics[width=0.35\textwidth]{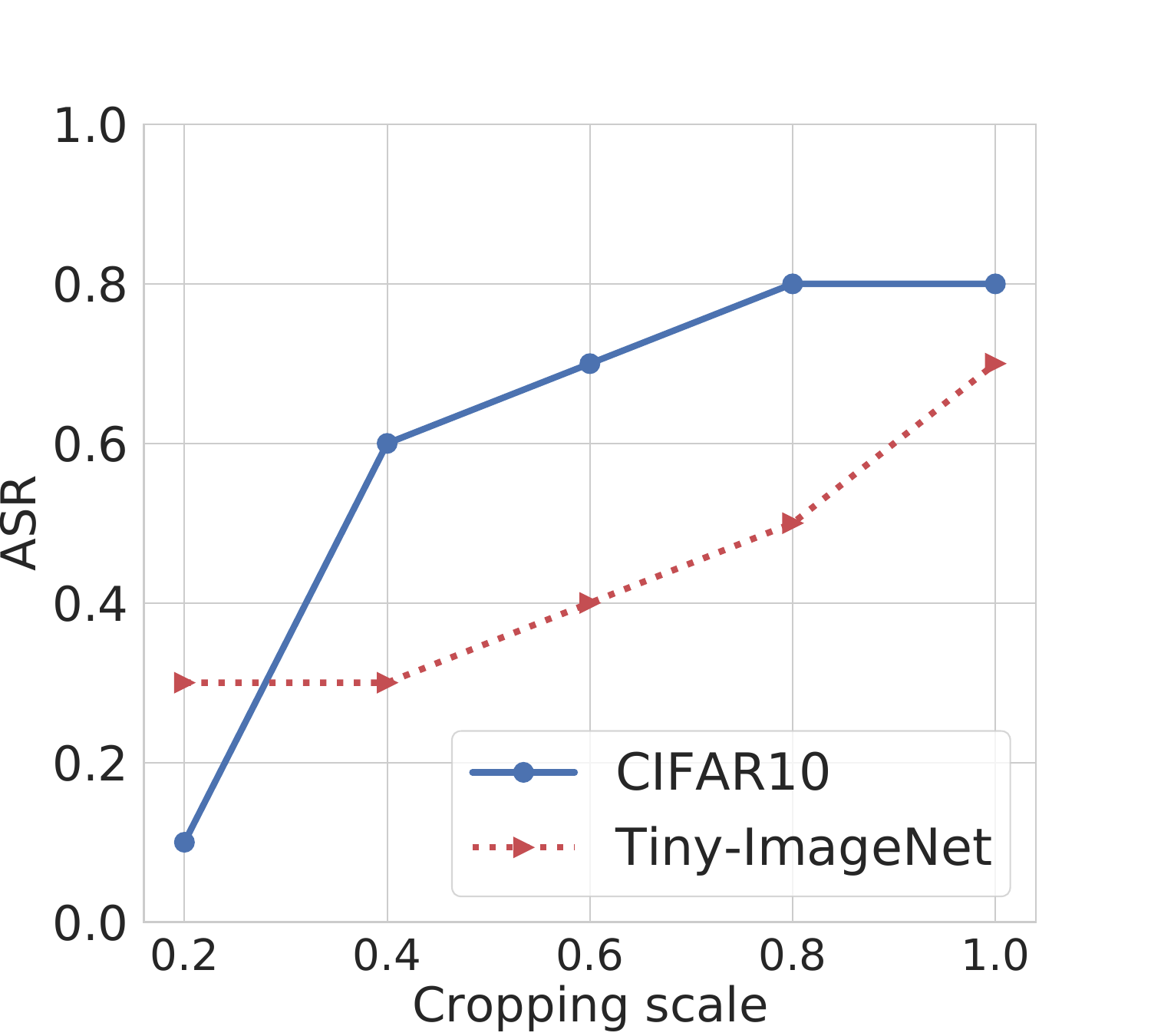} }
    \caption{ASR of PoisonedEncoder when  the attacker randomly crops a target/reference input with different cropping scales before using them to construct poisoning inputs, \SHF{where no defenses are deployed}.}
    \label{figure-random-crop}
\end{figure}

Defenses against data poisoning attacks can be categorized into \emph{pre-processing}, \emph{in-processing}, and \emph{post-processing} (details are discussed in Section~\ref{relatedwork-defense}). We explore one {pre-processing} defense (i.e., detecting poisoning inputs before pre-training), three {in-processing} defenses (i.e., re-designing the contrastive learning algorithm to be more robust), and one {post-processing} defense (i.e., fine-tuning a potentially poisoned encoder to remove the attack effect). In the following defense experiments, unless otherwise mentioned, we use the default parameter settings in Section~\ref{exp-setup} and consider STL10 as the target downstream task. 

\subsection{Pre-processing Defense}
\label{preprocessingdefense}
When there are a small number of reference inputs, \SHF{PoisonedEncoder} combines target inputs with a limited number of reference inputs to construct poisoning inputs. As a result, there are duplicate poisoning inputs. 
 In particular, the poisoning input crafted from one of the four combinations of a target input and a reference input may appear multiple times in the poisoned pre-training data. Therefore, the encoder provider can first remove duplicates in the pre-training data. However, such duplicates checking  is insufficient when the attacker has a large number of reference inputs. 
 Therefore, we further explore a clustering-based method to detect poisoning inputs after removing duplicate pre-training inputs. Our intuition is that the poisoning inputs contain  the same target inputs and thus may appear in the same clusters after clustering the pre-training inputs. Specifically, we group the pre-training images into $K$ clusters using $K$-Means with $\ell_2$-distance metric on the pixel values and predict the four clusters with the smallest average pair-wise $\ell_2$-distance as poisoning inputs. We consider the four clusters because  PoisonedEncoder uses the four combinations to construct poisoning inputs, each of which may correspond to a cluster.

 To evade such pre-processing defense, the attacker can randomly crop a target/reference input before combining them to construct a poisoning input. Specifically, the attacker crops a random square region of the target input and a random square region of the reference input with a \emph{cropping scale}, which is the ratio between the square region size and the target/reference input size. Then, the two random square regions are combined and resized to construct a poisoning input.  
 Figure~\ref{figure-random-crop} shows the ASR of PoisonedEncoder when the attacker randomly crops a target/reference input with different cropping scales when the pre-processing defense is not deployed, where cropping scale = 1 means no cropping. Our results show that ASR decreases as the cropping scale decreases when the defense is not deployed. This is because a smaller cropping scale makes it harder for the poisoned encoder to learn similar feature vectors for the original target input and reference inputs. 
 
\begin{figure}[!t]
    \centering
    \subfloat[Pre-trained on CIFAR10]
    {
      \includegraphics[width=0.23\textwidth]{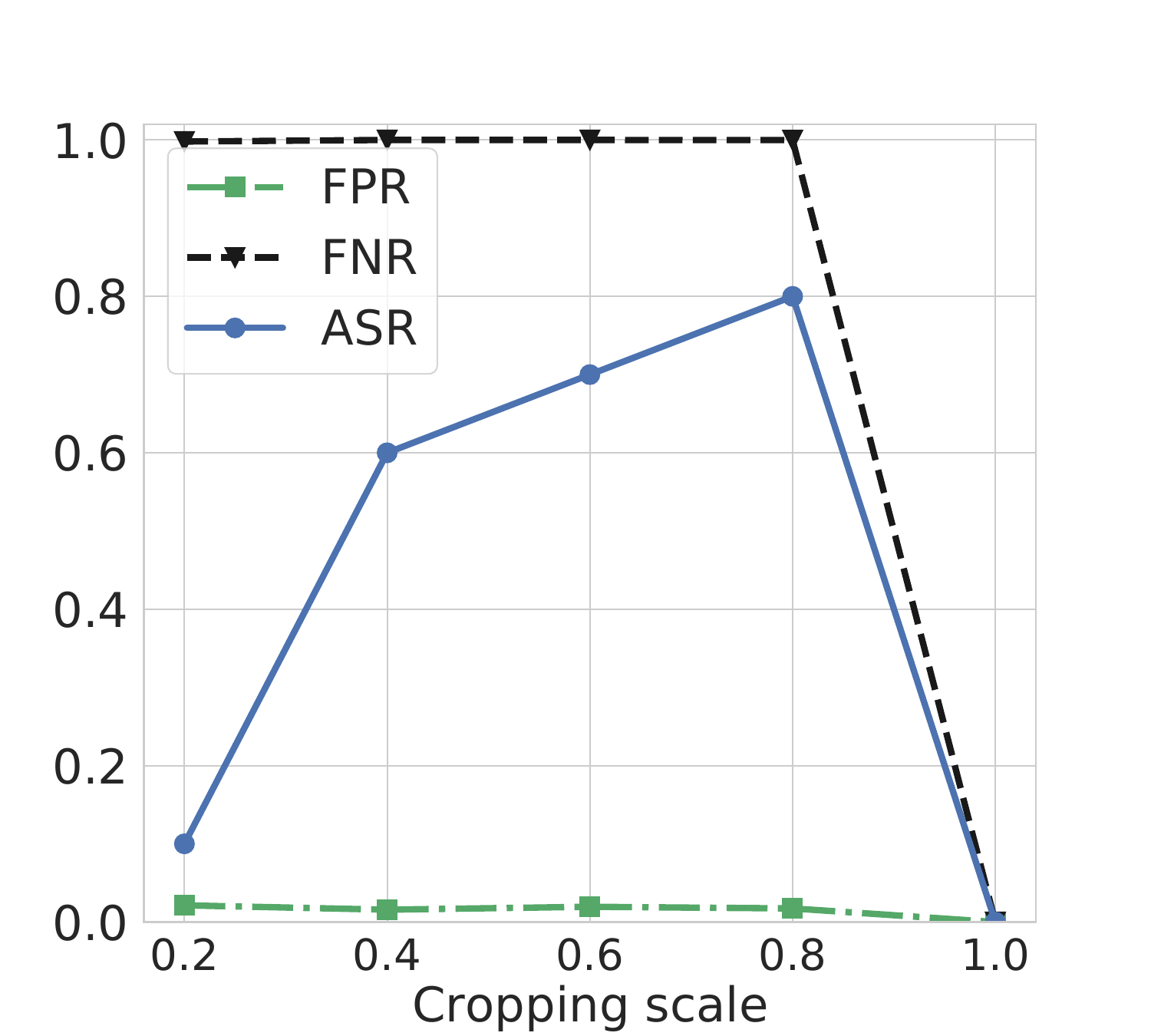}
    }
    \subfloat[Pre-trained on Tiny-ImageNet]{ 
    \includegraphics[width=0.23\textwidth]{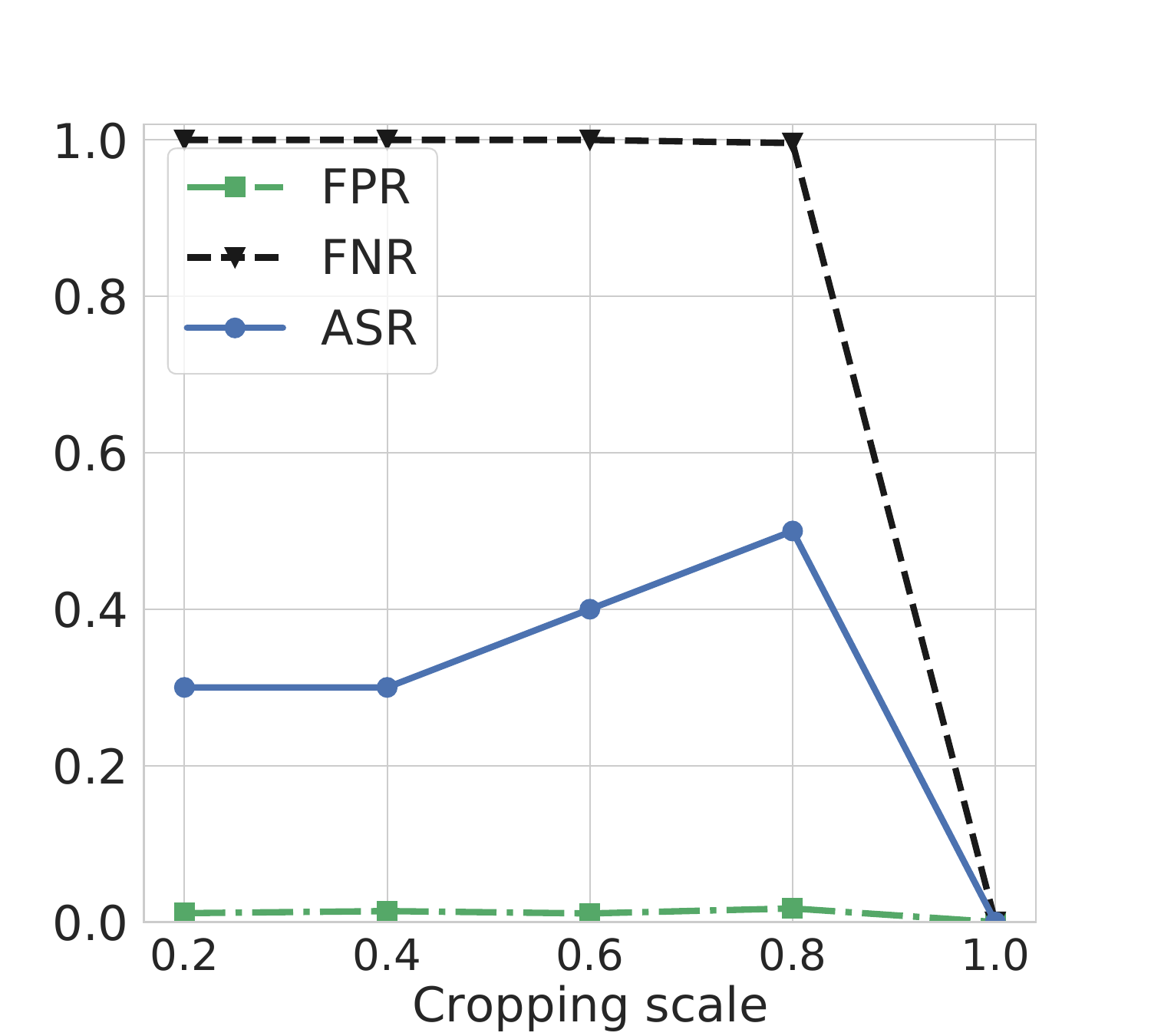}
    }
    \caption{Results of pre-processing defense. \SHF{The cropping scale refers to that an attacker crops a target/reference input before  using them to construct poisoning inputs.}} 
        \label{figure-defense-kmeans}
\end{figure}

 Figure~\ref{figure-defense-kmeans} shows the \emph{False Positive Rate (FPR)} and \emph{False Negative Rate (FNR)} of detecting poisoning inputs, as well as the ASR of PoisonedEncoder after removing the detected poisoning inputs and pre-training on the remaining inputs, where  $K=100$ for $K$-Means, and we treat poisoning input as ``positive'' and clean input as ``negative''. 
 When \SHF{PoisonedEncoder} does not use random cropping (i.e., cropping scale is 1) before combinations, the defense can effectively defend against PoisonedEncoder. However, the defense is ineffective, i.e., FNR is close to 1, when PoisonedEncoder randomly crops target/reference inputs to construct  poisoning inputs. For instance, PoisonedEncoder with cropping scale 0.8 still achieves high ASRs.

\subsection{In-processing Defenses}
\label{inprocessingdefense}
\begin{figure}[!t]
    \centering
    \subfloat[Pre-trained on CIFAR10 ]
    {\label{fig-impact-pre-training-epochs}
      \includegraphics[width=0.23\textwidth]{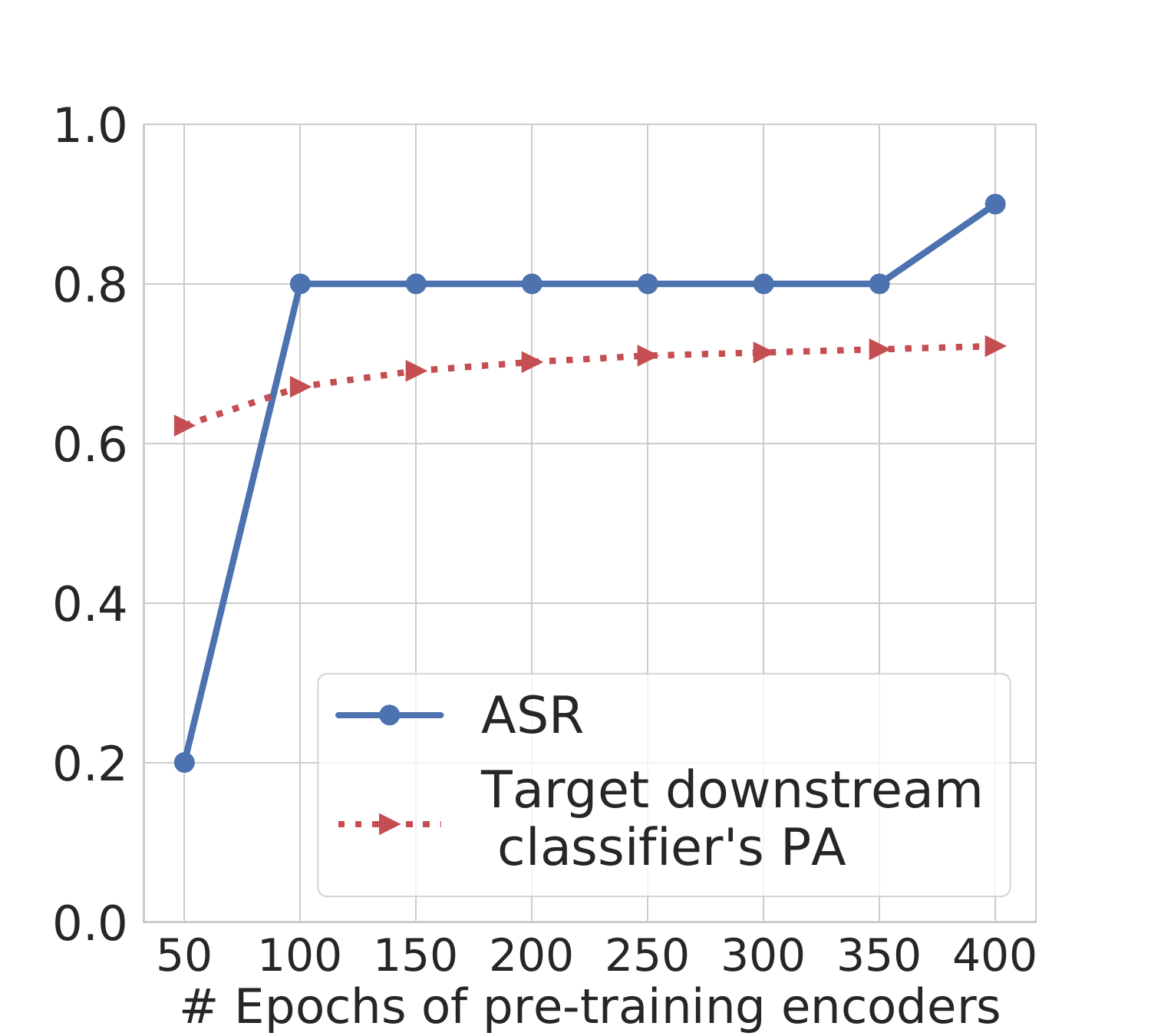}}
    \subfloat[Pre-trained on Tiny-ImageNet]{ {\label{fig-impact-pre-training-epochs-pre-tiny}}
    \includegraphics[width=0.23\textwidth]{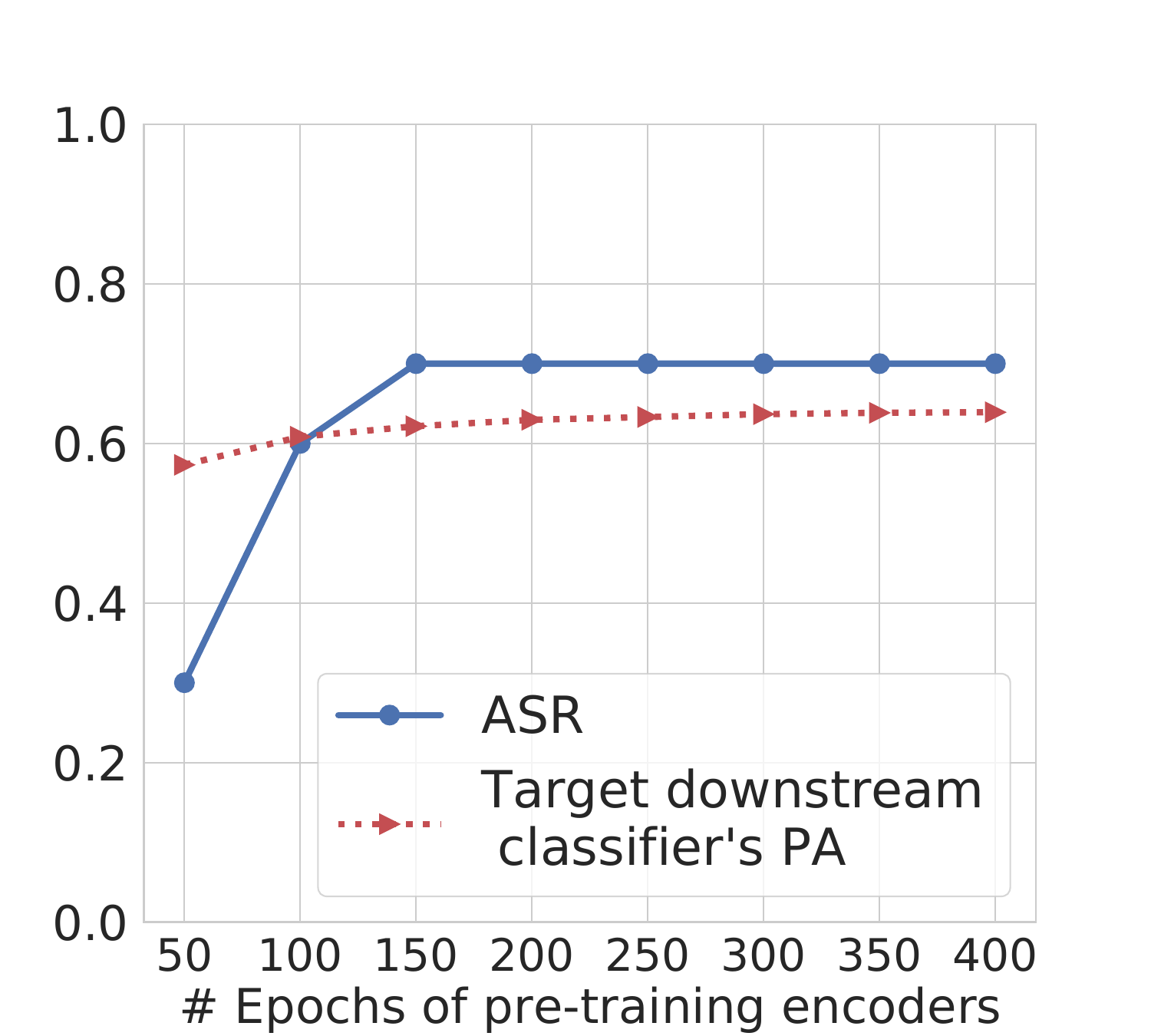}
    }
    
    \subfloat[Pre-trained on CIFAR10]
    {\label{fig-impact-fine-tuning-epochs}
      \includegraphics[width=0.23\textwidth]{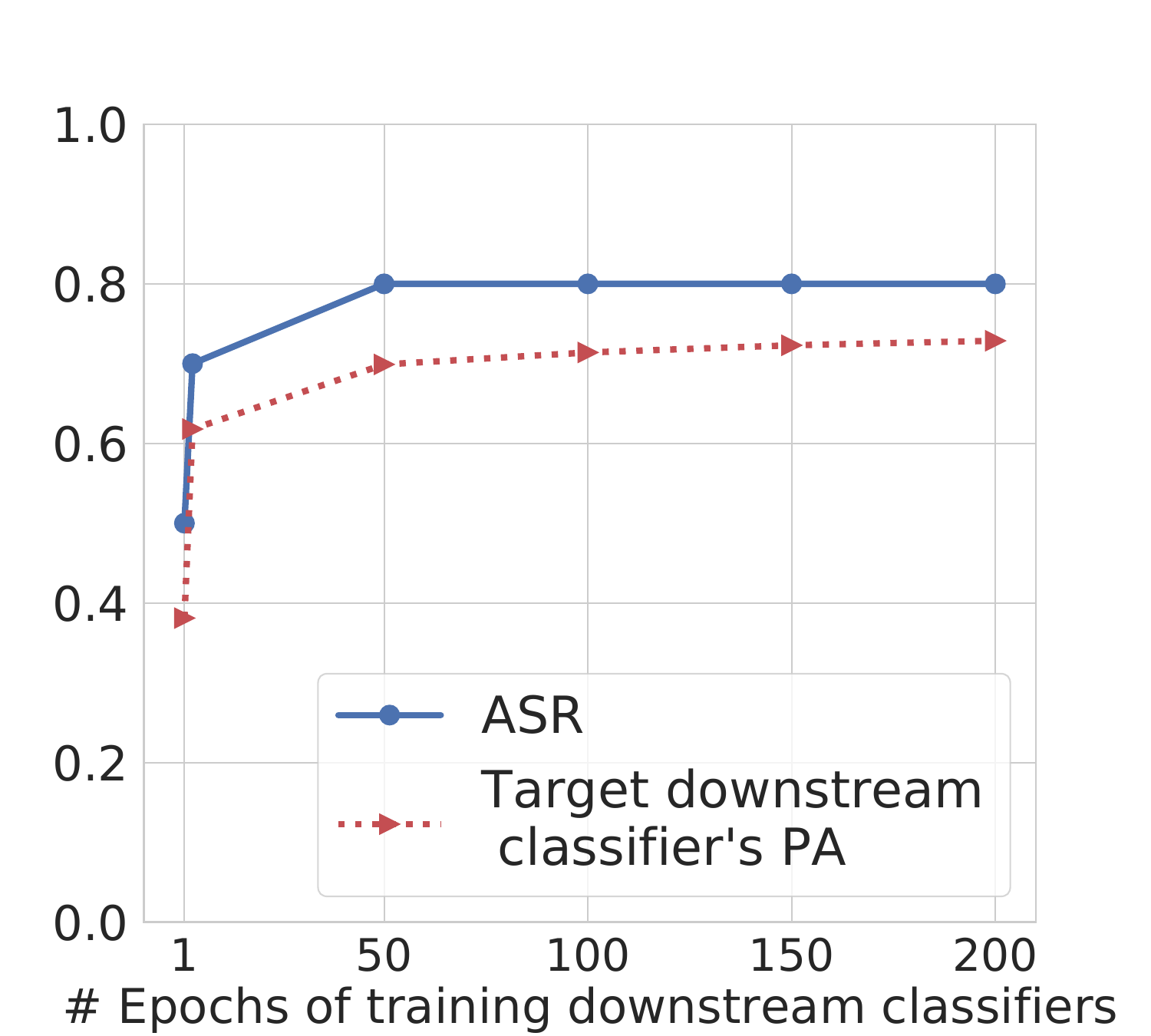}
    }
    \subfloat[Pre-trained on Tiny-ImageNet]{ \label{fig-impact-fine-tuning-epochs-pre-tiny}
    \includegraphics[width=0.23\textwidth]{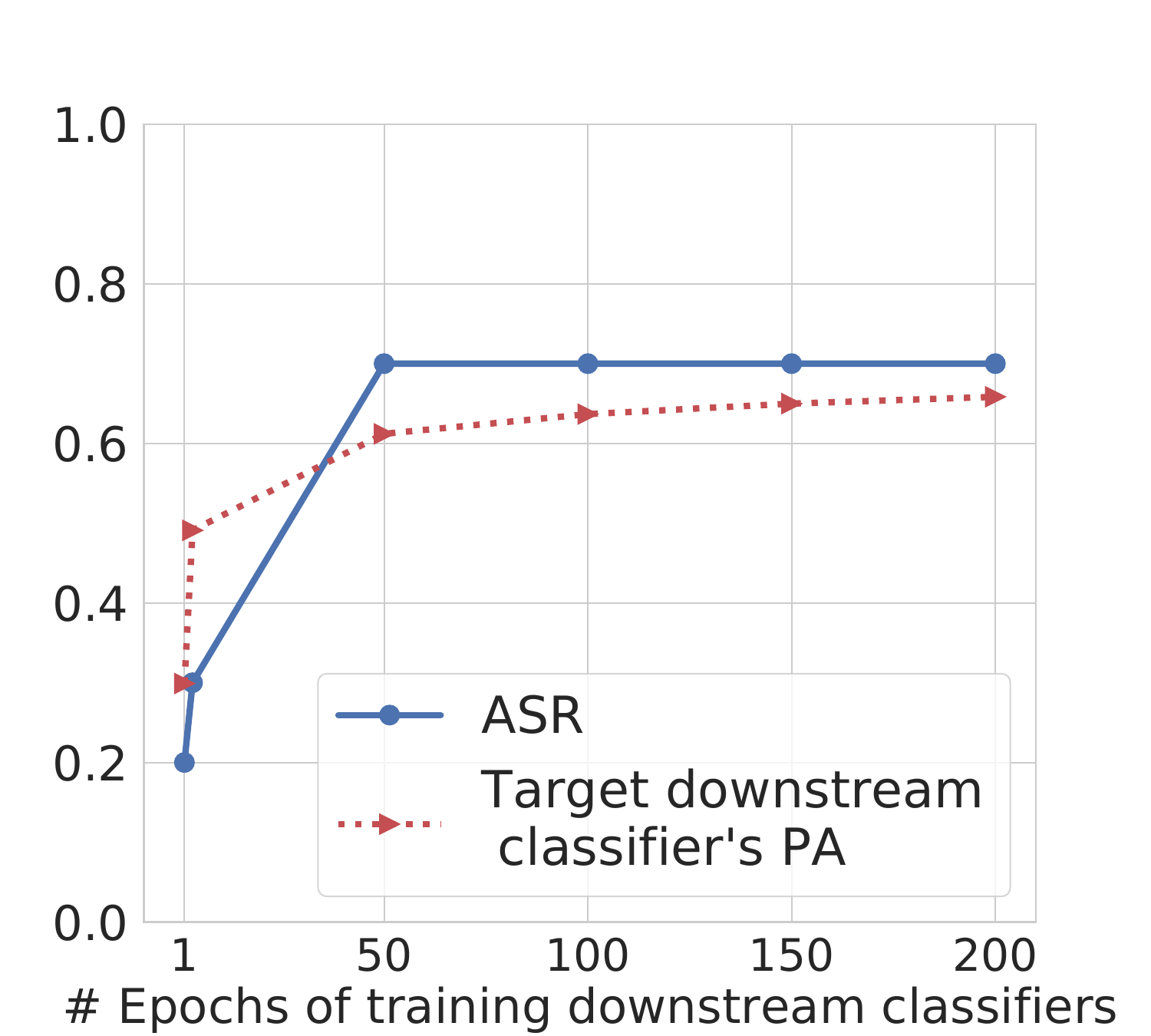}
    }
    \caption{In-processing defense: early stopping. First row: early stopping when pre-training an encoder. Second row: early stopping when training a downstream classifier.} 
    \label{figure-defense-pre}
\end{figure}

\myparatight{Early stopping} Intuitively, PoisonedEncoder relies on enough pre-training epochs to make the poisoned encoder produce similar feature vectors for a target input and the reference inputs, and enough training epochs of the target downstream classifier to predict the reference/target inputs as the target class. Therefore, early stopping, in which an encoder is pre-trained or a downstream classifier is trained  using less epochs, could mitigate PoisonedEncoder. 
Figure~\ref{figure-defense-pre} shows the ASR of PoisonedEncoder and the target downstream classifier's PA as a function of the number epochs used to pre-train an encoder or train the target downstream classifier.  We observe that early stopping can reduce ASR when pre-training an encoder or training a downstream classifier using less epochs. However, such early stopping also reduces the accuracy of the downstream classifiers built based on the poisoned encoder. To summarize, early stopping can mitigate PoisonedEncoder at the cost of sacrificing utility.

\myparatight{Bagging} Jia et al.~\cite{jia2020intrinsic} showed that bagging, a well-known ensemble method, has intrinsic certified robustness against data poisoning attacks. The idea is to create multiple random subsamples of training data and learn a base classifier on each subsample; and we take majority vote among the base classifiers to classify a testing input.  We extend bagging to contrastive learning as a defense against PoisonedEncoder. In particular, we create multiple random subsamples of pre-training data and pre-train a \emph{base encoder} on each subsample using SimCLR; given a downstream task, we train a base downstream classifier based on each base encoder; and we take majority vote among the base downstream classifiers to classify a testing input. Such bagging-based method provably predicts the same label for a testing input when the number of poisoning inputs in the pre-training dataset is bounded.

Table~\ref{table_defense_bagging} shows the ASR of PoisonedEncoder and the PA of the target downstream task, where each subsample includes 500 randomly selected pre-training inputs and we create 100 subsamples/base encoders/base classifiers.
The ``Dropped Ratio'' in the table is the decreased percentage of PA in Table~\ref{table_defense_bagging}, compared to that in Table~\ref{table_accuracy}. 
Our results show that bagging can defend against PoisonedEncoder at the cost of substantially reducing the accuracy of the downstream classifiers. 

\begin{table}[!t]\renewcommand{\arraystretch}{1.3} 
	\fontsize{7.5}{8}\selectfont
	\centering
	\caption{In-processing defense: bagging.  ``Dropped Ratio'' is the decreased percentage of the PA when bagging is used, compared to  when bagging is not used. }
	\setlength{\tabcolsep}{1mm}
	{
	\begin{tabular}{|c|c|c|c|}
		\hline
	\makecell{Pre-training Dataset} & \makecell{ASR}  & \makecell{PA} & \makecell{Dropped Ratio (\%)} \\ \hline
	CIFAR10 & 0.0         &  0.431    & 39.2\\
	\hline
	Tiny-ImageNet & 0.0   & 0.415 & 34.9  \\
	\hline
	\end{tabular}
	}
	\label{table_defense_bagging}
\end{table}

\myparatight{Pre-training without random cropping} PoisonedEncoder exploits the random cropping data augmentation operation, which is widely used in  contrastive learning. Therefore, a countermeasure against PoisonedEncoder is to not use random cropping during pre-training.  Table~\ref{table_defense_no_crop} shows the ASR of PoisonedEncoder and the PA of the target downstream task when random cropping is not used by SimCLR for pre-training.  Our results show that pre-training without random cropping can mitigate PoisonedEncoder, but it sacrifices the utility of the encoder substantially. Our observation is consistent with Chen et al.~\cite{chen2020simple}, which showed that random cropping is crucial for contrastive learning.

\begin{table}[!t]\renewcommand{\arraystretch}{1.3} 
	\fontsize{7.5}{8}\selectfont
	\centering
	\caption{In-processing defense: pre-training without random cropping. ``Dropped Ratio'' is the decreased percentage of the PA when  random cropping is not used, compared to when random cropping is used.}
	\setlength{\tabcolsep}{1mm}
	{
	\begin{tabular}{|c|c|c|c|}
		\hline
	\makecell{Pre-training Dataset} & \makecell{ASR} &  \makecell{PA} & \makecell{Dropped Ratio (\%)} \\ \hline
	CIFAR10 & 0.1 & 0.391    & 45.3\\
	\hline
	Tiny-ImageNet & 0.1 & 0.380 & 40.3  \\
	\hline
	\end{tabular}
	}
	\label{table_defense_no_crop}
	\vspace{-2mm}
\end{table}

\subsection{Post-processing Defense}

\begin{figure}[!t]
    \centering
    \subfloat[Pre-trained on CIFAR10 ]
    {
      \includegraphics[width=0.23\textwidth]{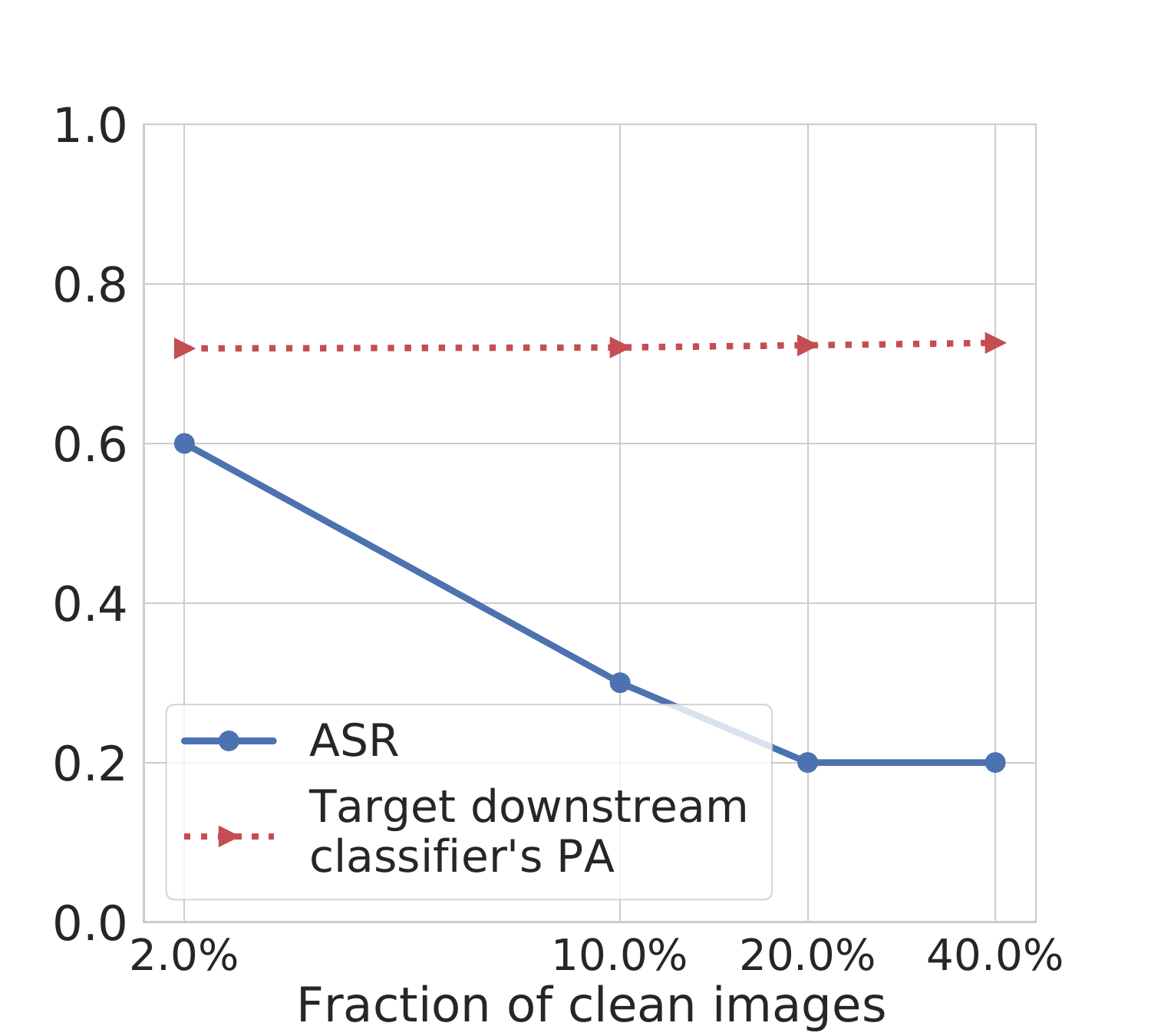}
    }
    \subfloat[Pre-trained on Tiny-ImageNet]{ 
    \includegraphics[width=0.23\textwidth]{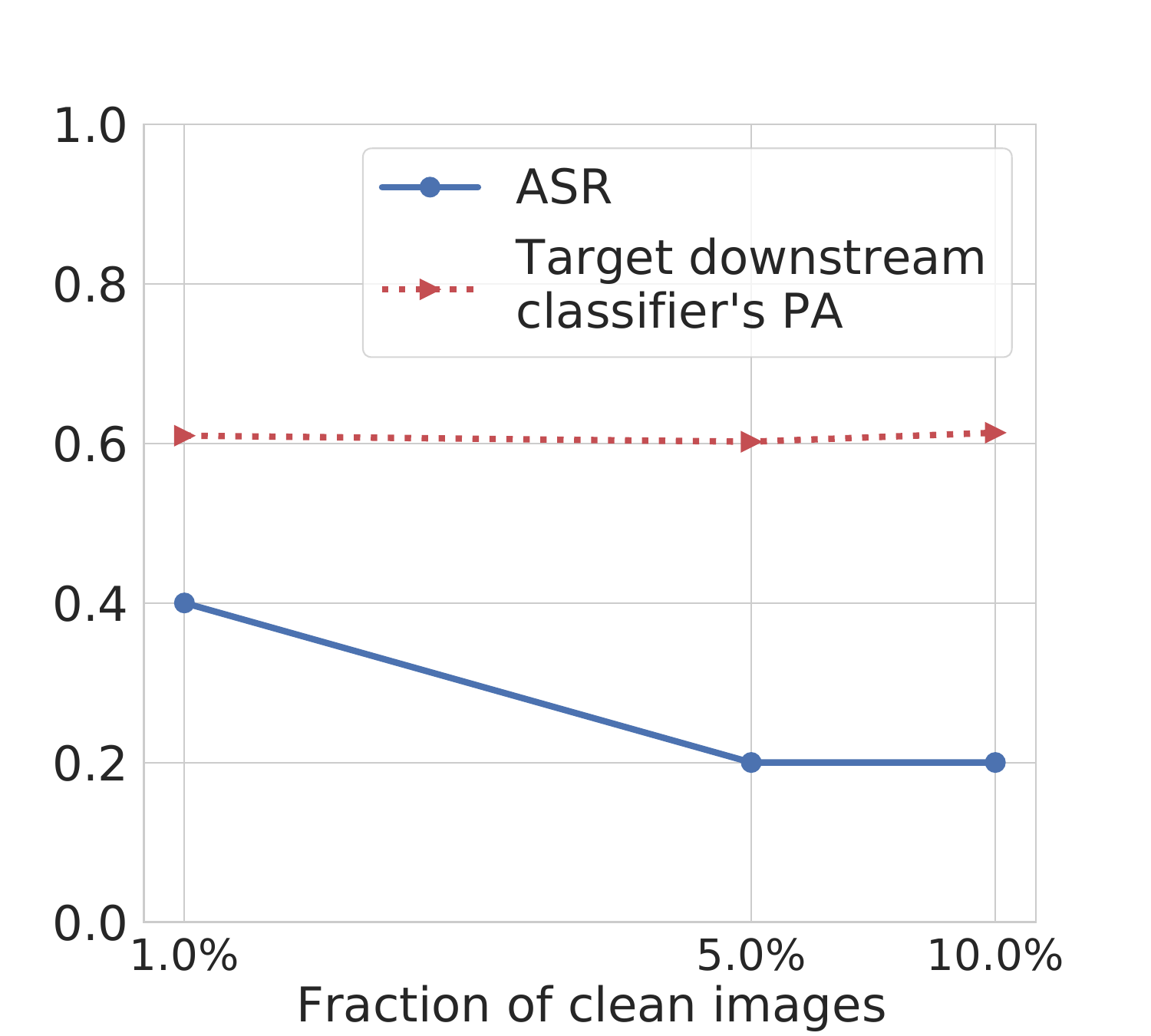}
    }
    \caption{Post-processing defense. The x-axis is in log-scale.}
        \label{figure-defense-post}
\end{figure}

In this defense, the encoder provider aims to remove the attack effect from a potentially poisoned encoder by fine-tuning it  for extra epochs on some clean images. Figure~\ref{figure-defense-post} shows the ASR of PoisonedEncoder and the PA of the target downstream task when we randomly sample a certain fraction of clean images in the pre-training dataset for fine-tuning, where the number of fine-tuning epochs is 300 and the fine-tuning learning rate is 0.0001. Our results show that fine-tuning can reduce the ASR of PoisonedEncoder without sacrificing the encoder's utility. However, it requires manually collecting a large set of clean images, which is time-consuming.  

\section{Related Work}

\subsection{Data Poisoning Attacks}
Data poisoning attacks generally refer to tampering with the training data of a machine learning system such that the poisoned model makes incorrect predictions as an attacker desires~\cite{barreno2006can}. Specifically, the attacker may desire the model to make incorrect predictions for indiscriminate testing inputs (i.e., have low testing accuracy), which are known as \emph{untargeted data poisoning attacks}, or desire the model to make attacker-chosen, incorrect predictions for attacker-chosen testing inputs, which are known as \emph{targeted data poisoning attacks}. We study targeted data poisoning attacks to contrastive learning in this work.

\myparatight{Data poisoning attacks to supervised learning}
Data poisoning attacks have been studied extensively for various supervised learning algorithms such as 
support vector machine~\cite{biggio2012poisoning}, logistic regression~\cite{wen2020palor}, and neural networks~\cite{munoz2017towards,suciu2018does,shafahi2018poison,geiping2020witches,jagielski2021subpopulation}. 
These attacks tamper with the features and/or labels of training examples to perform untargeted data poisoning attacks~\cite{biggio2012poisoning,munoz2017towards} or targeted data poisoning attacks~\cite{shafahi2018poison,geiping2020witches,jagielski2021subpopulation,suciu2018does}. These attacks are often  formulated as bilevel optimization problems~\cite{biggio2012poisoning,steinhardt2017certified}, where the attacker's objective on the poisoned model is formulated as the outer optimization while learning the poisoned model on the poisoned training data is formulated as the inner optimization. Data poisoning attacks to different machine learning algorithms instantiate different bilevel optimization problems, which are often challenging to solve and require customized, heuristic solutions. We extended a state-of-the-art targeted data poisoning attack~\cite{geiping2020witches} to neural network based classifier to contrastive learning; and our results show that such extended attack achieves suboptimal attack success rate due to the challenge of solving our formulated bilevel optimization problem.

\myparatight{Data poisoning attacks to semi-supervised learning} Different from supervised learning, semi-supervised learning uses both labeled and unlabeled training data. Therefore, other than tampering with the labeled training data, an attacker can also tamper with the unlabeled training data to attack semi-supervised learning. For instance, Wang and Gong~\cite{wang2019attacking} proposed data poisoning attacks to graph-based semi-supervised learning methods via manipulating the graph structure. Xu et al.~\cite{xu2020attacking} extended such attack by adding new fake nodes without manipulating existing graph structure. Carlini~\cite{carlini2021poisoning} proposed a targeted data poisoning attack to  semi-supervised learning for image classification, which tampers with the unlabeled training data. Roughly speaking, he proposed a heuristic approach to craft unlabeled poisoning inputs that are interpolations between a target input and reference inputs. We extended this attack to contrastive learning  and our results show that such extended attack achieves suboptimal attack success rate due to the difference between semi-supervised learning and contrastive learning.

\myparatight{Data poisoning attacks to contrastive learning} Data poisoning attacks to contrastive learning are much less explored. To the best of our knowledge, Carlini and Terzis~\cite{carlini2021poisoning_clip} is the only work on data poisoning attacks to contrastive learning. However, they focus on multi-modal contrastive learning, which pre-trains encoders on (image, text) pairs.
In particular, their attack constructs text captions containing the attacker-chosen target class and associates them with the attacker-chosen target image to generate the poisoning (target image, text captions) pairs, which are injected into the pre-training dataset.  We note that their attack is not applicable to image-only contrastive learning, which is the focus of our work,  because the images are not associated with text.

\myparatight{Other data poisoning attacks}
Data poisoning attacks have been proposed to many other machine learning systems. Examples include recommender systems~\cite{li2016data,yang2017fake,fang2018poisoning,fang2020influence,huang2021data},  federated analytics~\cite{bhagoji2019analyzing,fang2020local,ma2019data,cao2021data,wu2021poisoning,fang2021data,cao2022mpaf}, search engines~\cite{joslin2019measuring}, as well as natural language models~\cite{schuster2020humpty,schuster2021you}. For instance, in federated analytics, clients perform computation on their data locally and send computation results instead of raw data to a cloud server. As a result, other than tampering with their data, malicious clients (fake clients or compromised genuine ones) can also tamper with the computation process, which result in stronger poisoning attacks~\cite{fang2020local,cao2021data}. 

\myparatight{Backdoor attacks} Backdoor attacks~\cite{gu2017badnets,chen2017targeted,liu2017trojaning,jia2021badencoder} also tamper with the training phase, e.g., an attacker embeds a trigger to some training inputs and changes their labels to an attacker-chosen one. A model learnt on such poisoned training data predicts the attacker-chosen label for any input once the attacker embeds the trigger into it. Unlike data poisoning attacks that tamper with the training phase alone, backdoor attacks tamper with both training and testing phases.

\subsection{Defenses against Data Poisoning Attacks}
\label{relatedwork-defense}
Depending on the stage of a machine learning pipeline where a defense is deployed, we can categorize defenses against data poisoning attacks into \emph{pre-processing defenses}, \emph{in-processing defenses}, and \emph{post-processing defenses}. Pre-processing defenses aim to detect and remove poisoning training examples before the training process starts; in-processing defenses aim to re-design the training algorithm such that it can learn an accurate and clean model even if some training examples are poisoned; and post-processing defenses aim to remove the attack effect from a model that has already been trained on (potentially) poisoned training data.

\myparatight{Pre-processing defenses}  Some defenses~\cite{paudice2018label,peri2020deep} in this category detect poisoning training examples based on the label-mismatch between  a training example and its nearest neighbors. In particular, Paudice et al.~\cite{paudice2018label} proposed to relabel a training input as the most frequent label among its $k$-nearest neighbors, and the relabeled training data are used to train a classifier. Peri et al.~\cite{peri2020deep} proposed to remove  training examples whose labels are not the most frequent ones among their $k$-nearest neighbors before training a classifier. These defenses are limited to supervised learning as they rely on labels of the training examples, and thus are not applicable to contrastive learning.   
Other pre-processing defenses aim to detect poisoning training examples based on anomaly detection~\cite{cretu2008casting,paudice2018detection,chen2021pois}, which can be broken by strong, adaptive attacks~\cite{koh2018stronger}. In our work, we tailor an anomaly detection based pre-processing defense to counter PoisonedEncoder, but our results show that such defense is ineffective when  PoisonedEncoder randomly crops a target/reference input before using them to construct poisoning inputs.

\myparatight{In-processing defenses} Some in-processing defenses~\cite{barreno2010security,suciu2018does,tran2018spectral}  train an initial model using the potentially poisoned training data,  remove potentially poisoned training examples based on the initial model, and re-train a model using the remaining training examples. For instance, \emph{Reject on Negative Impact (RONI)}~\cite{barreno2010security}  measures the impact of each training example on the error rate of the initial model and removes the training examples that have large negative impact. Tran et al.~\cite{tran2018spectral} removes a certain fraction of the most abnormal training examples, where the abnormality of a training example is measured by the spectral property of its feature representation obtained from the initial model.
 Carlini~\cite{carlini2021poisoning} proposed an in-processing defense that is tailored to their poisoning attack to the unlabeled training data in semi-supervised learning. Their key assumption is that the predicted labels of benign unlabeled training examples are influenced by many other unlabeled examples simultaneously, while the predicted labels of poisoning unlabeled examples are predominately influenced by other poisoning examples. Based on this assumption, their defense calculates a score for each unlabeled training example based on the influence of its nearest neighbors on its labels predicted by the initial model during training; and the scores are used to detect and remove poisoned training examples.

These defenses 1) are not applicable to contrastive learning~\cite{barreno2010security,suciu2018does,carlini2021poisoning}, e.g., it is challenging to define ``error rate'' of  an encoder for a pre-training input and encoder does not predict labels for the unlabeled pre-training inputs, or 2) require information that may not be available to a defender~\cite{tran2018spectral}, e.g., an upper bound of the fraction of poisoned training examples.

All the defenses above do not have certified robustness guarantees. 
Some works~\cite{wang2020certifying,jia2020certified,rosenfeld2020certified,wang2021certified} propose new learning algorithms that have certified robustness, while some works~\cite{jia2020intrinsic,jia2022certified} analyze the intrinsic certified robustness of existing learning algorithms. A learning algorithm is certifiably robust against data poisoning attacks if its learnt classifier provably predicts the same label for a testing input when the number of poisoned training examples is bounded. For instance, Jia et al.~\cite{jia2020certified} and Wang et al.~\cite{wang2021certified}  leveraged randomized smoothing to build graph-based algorithms that are certifiably robust  against graph-structure poisoning. Jia et al.~\cite{jia2020intrinsic,jia2022certified} showed that bagging and nearest neighbors have intrinsic certified robustness against data poisoning attacks. We extend bagging to contrastive learning and our results show that it can defend against PoisonedEncoder but sacrifices utility substantially. 

We also propose two in-processing defenses (early stopping and no random cropping) that are tailored to contrastive learning, but our results show that they also sacrifice utility of the encoder.

\myparatight{Post-processing defenses} 
Several studies~\cite{liu2018fine,wang2019neural} proposed to post-process a potentially poisoned classifier to remove the attack effect. These methods often require a clean training dataset. For instance,  a potentially poisoned classifier can be fine-tuned using a clean training dataset~\cite{liu2018fine}. We extend fine-tuning to post-process an encoder as a defense against  PoisonedEncoder. Our results show that such defense can reduce the attack success rate of PoisonedEncoder but requires a clean pre-training dataset, which may be hard to collect.  
\section{Extending to Backdoor Attack}
\begin{figure}[!t]
    \centering
    \subfloat[]{
      \includegraphics[width=0.1\textwidth]{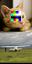}
    }
    \hspace{18mm} 
    \subfloat[]{
      \includegraphics[width=0.1\textwidth]{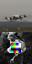}
    }
    
    \subfloat[]{
      \includegraphics[width=0.2\textwidth]{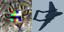}
    }
    \subfloat[]{
      \includegraphics[width=0.2\textwidth]{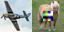}
    } 
    \caption{Examples of poisoning inputs when extending PoisonedEncoder to backdoor attack. The target downstream task is STL10 and the target class is `airplane'.}
    \label{fig:backdoor}
\end{figure}

Backdoor attack is another category of targeted poisoning attacks. 
In a backdoor attack, a target input is any input embedded with an attacker-chosen trigger. 
We can extend PoisonedEncoder to data poisoning based backdoor attack, where a downstream classifier built based on a poisoned encoder for a target downstream task predicts attacker-chosen target class for any trigger-embedded testing input. We assume the attacker has some inputs from the target class as reference inputs. Moreover, we assume the attacker has some inputs from the non-target classes as \emph{auxiliary inputs}. An attacker constructs poisoning inputs using the reference inputs and auxiliary inputs.  In particular, the attacker crafts a poisoning input as the concatenation of a random trigger-embedded auxiliary input (i.e., sample an auxiliary input and embed trigger into it) and a random reference input. Figure~\ref{fig:backdoor} illustrates several examples of poisoning
inputs  when the target downstream task is STL10 and the target class is `airplane'. In our experiments, we randomly sample 50 testing inputs from the target class `airplane' as reference inputs, randomly sample 500 testing inputs from the non-target classes as auxiliary inputs, and use poisoning rate of 1\%.  The pre-training dataset is CIFAR10 and the trigger is a 10 × 10 patch with random pixel values. Our attack is effective, e.g., the downstream classifier built based on the poisoned encoder for STL10 predicts the target class for 96.54\% of trigger-embedded testing inputs on average over 10 repeated experimental trials.

\section{Conclusion and Future Work}

In this work, we show that single-modal contrastive learning is vulnerable to targeted data poisoning attack.  An attacker can exploit the random cropping operation that contrastive learning relies on to attack  contrastive learning. In particular, an attacker combines a target input and reference inputs from the target class to construct poisoning inputs. Our evaluation shows that our attack is successful and maintains utility of the encoder. Moreover, extending existing targeted data poisoning attacks tailored to supervised learning and semi-supervised learning to contrastive learning achieves suboptimal attack success rates. Our evaluation on five defenses show that they are insufficient, i.e., they sacrifice the encoder's utility or require a large clean pre-training dataset. An interesting future work is to develop new defenses against our attack.

\SHF{\section*{Acknowledgements}

We thank our shepherd Nicholas Carlini and the anonymous reviewers for their constructive comments. This work was supported by NSF under Grant No. 2125977, 2112562, and 1937786 as well as the Army Research
Office under Grant No. W911NF2110182.}

\bibliographystyle{plain}
\bibliography{ccs-sample}
\appendix
\begin{figure}[!ht]
    \centering
    \subfloat[]
    {
      \includegraphics[width=0.16\textwidth, height =0.16\textwidth]{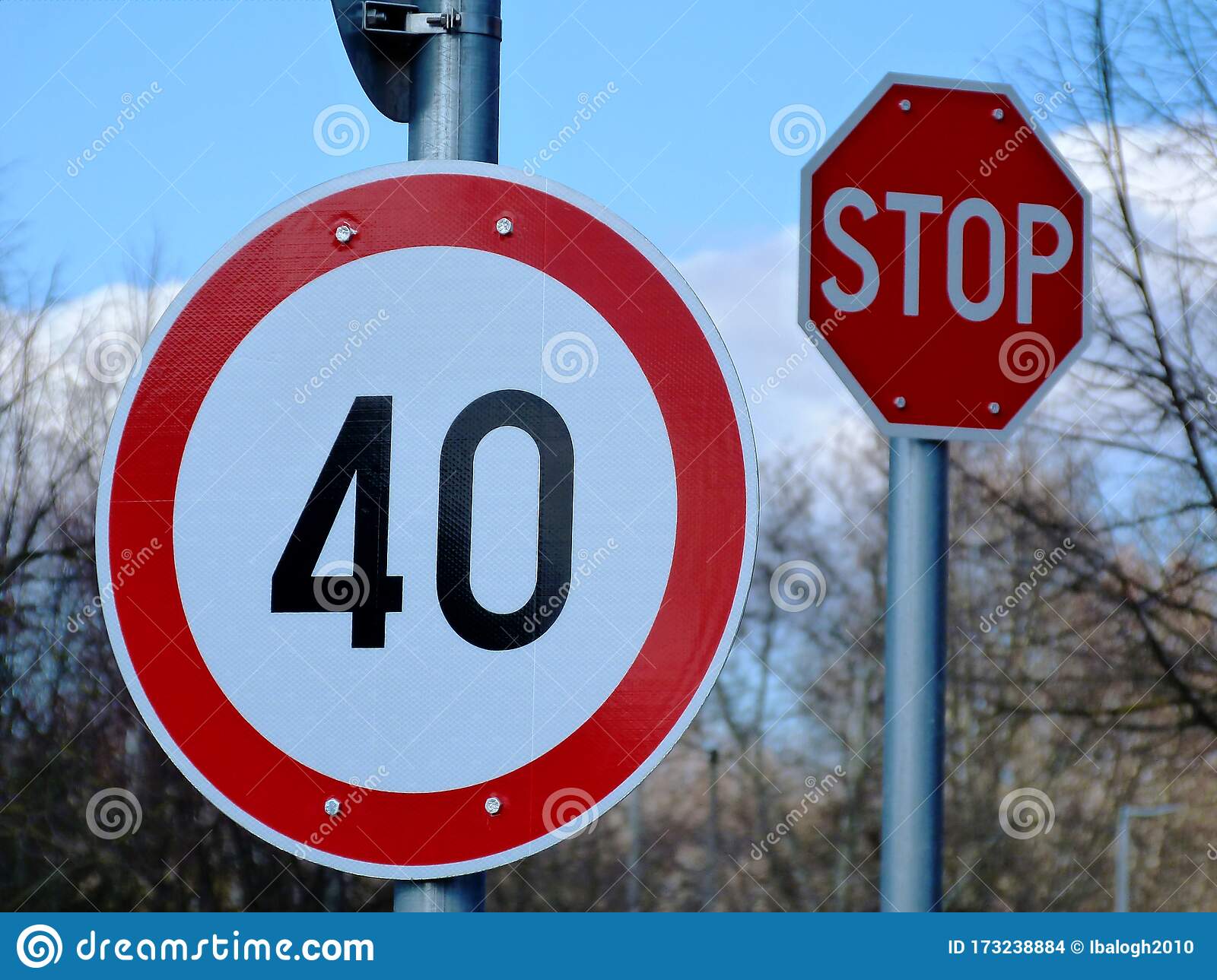}
    }
    \subfloat[]{ 
    \includegraphics[width=0.16\textwidth, height =0.16\textwidth]{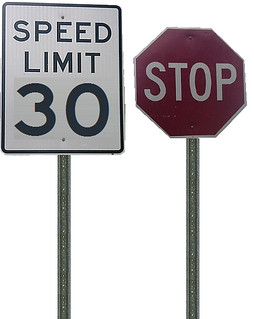}
    }
    \subfloat[]
    {
      \includegraphics[width=0.16\textwidth, height =0.16\textwidth]{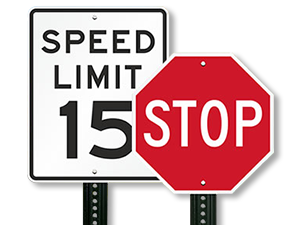}
    }\\
    \subfloat[]{ 
    \includegraphics[width=0.16\textwidth, height =0.16\textwidth]{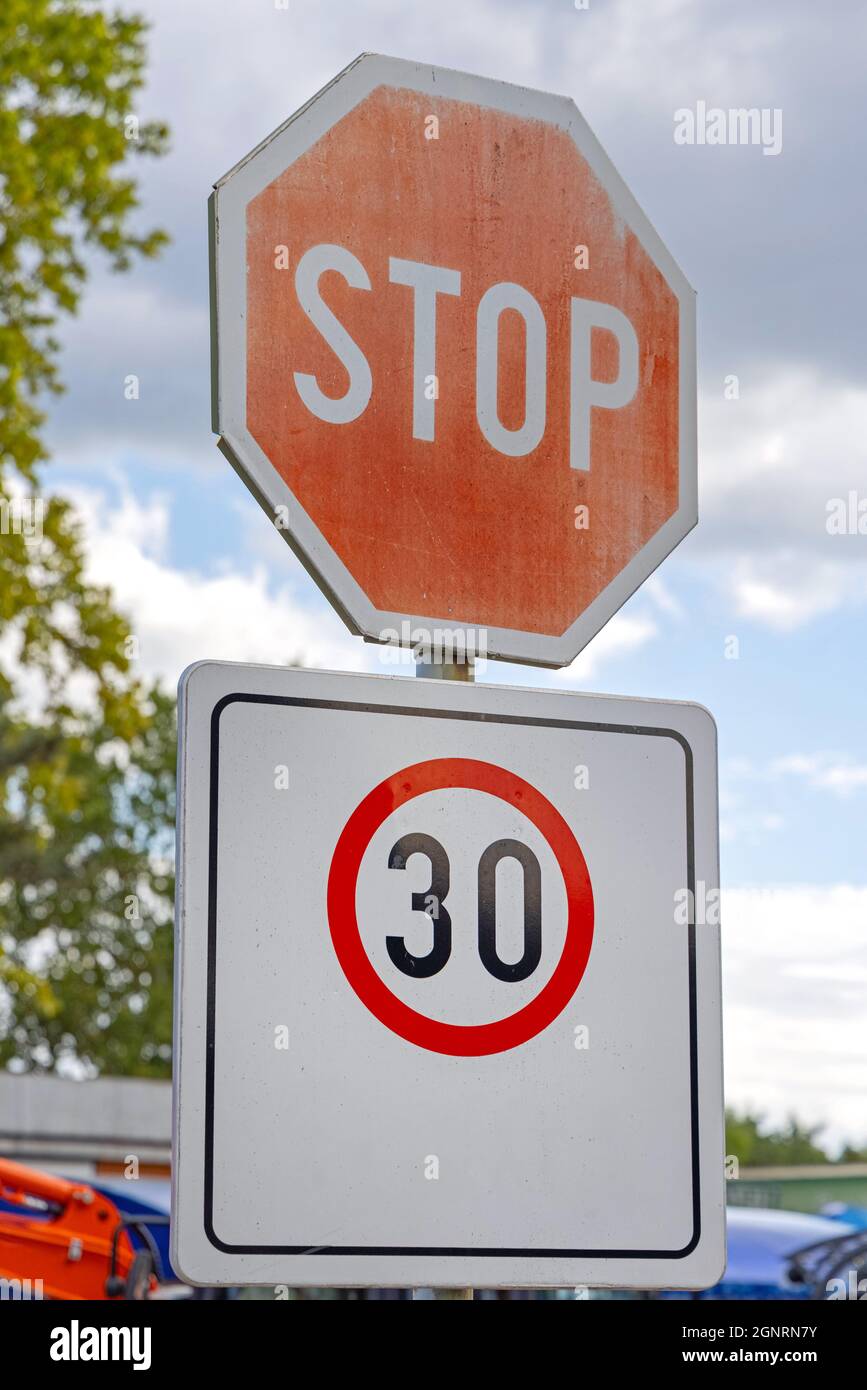}
    }
    \subfloat[]
    {
      \includegraphics[width=0.16\textwidth, height =0.16\textwidth]{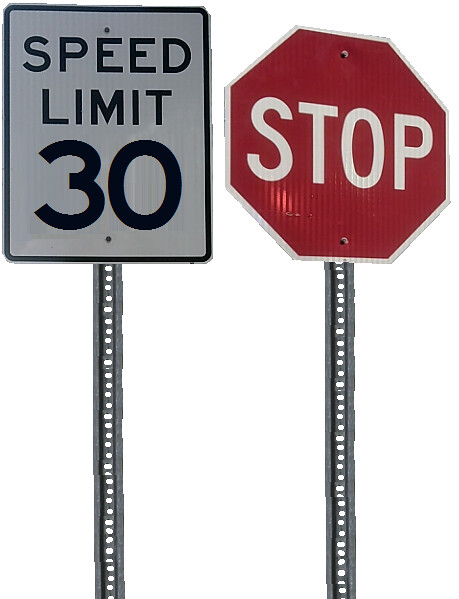}
    }
    \subfloat[]
    {
      \includegraphics[width=0.16\textwidth, height =0.16\textwidth]{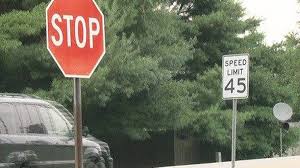}
    }\\
    \subfloat[]
    {
      \includegraphics[width=0.16\textwidth]{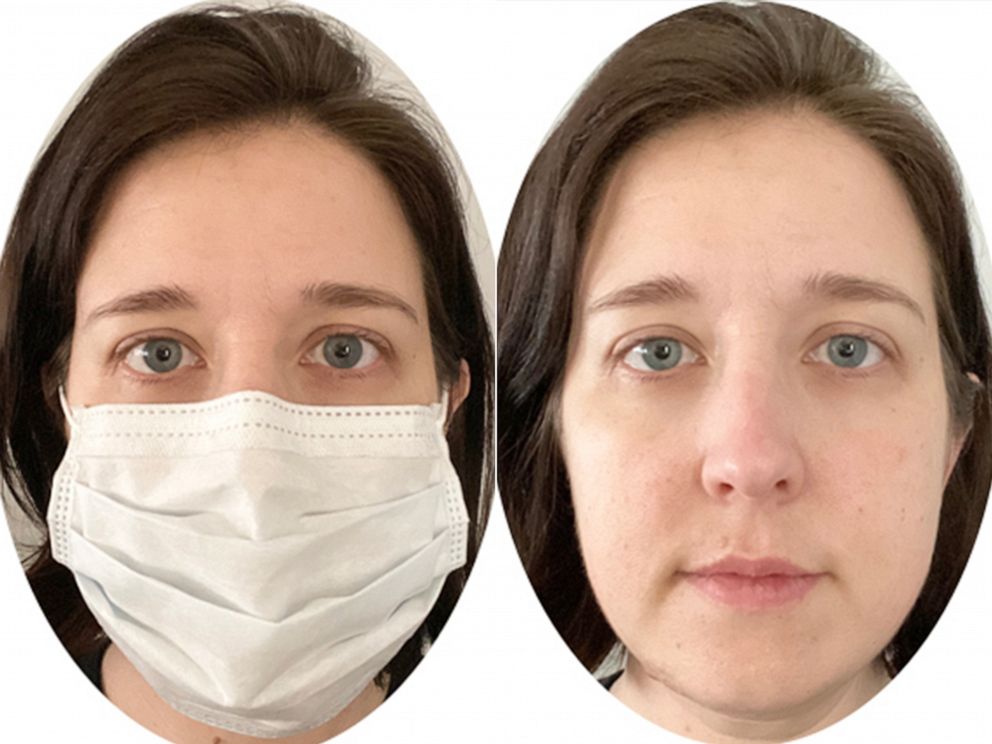}
    }
    \subfloat[]{ 
    \includegraphics[width=0.16\textwidth]{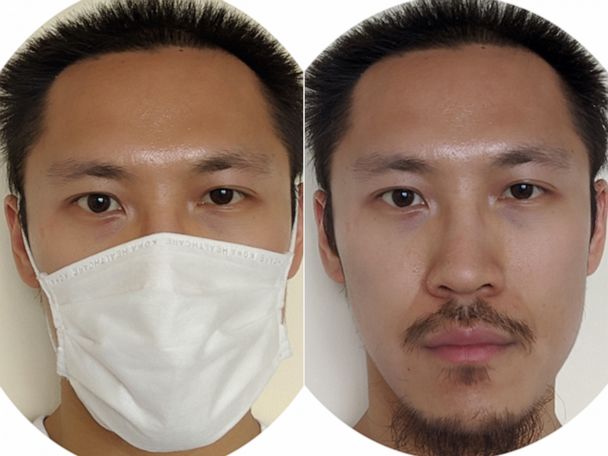}
    }
    \subfloat[]
    {
      \includegraphics[width=0.16\textwidth]{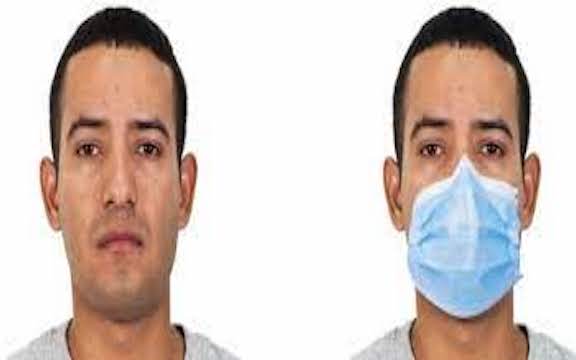}
    }\\

    \subfloat[]
    {
      \includegraphics[width=0.16\textwidth, height =0.16\textwidth]{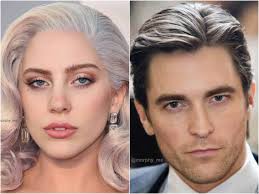}
    }
    \subfloat[]{ 
    \includegraphics[width=0.16\textwidth, height =0.16\textwidth]{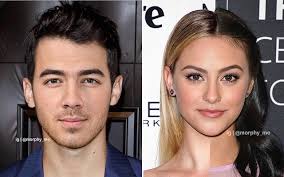}
    }
    \subfloat[]
    {
      \includegraphics[width=0.16\textwidth, height =0.16\textwidth]{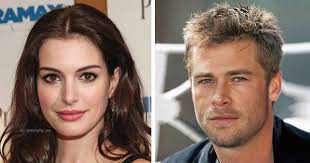}
    }\\
    \subfloat[]{ 
    \includegraphics[width=0.16\textwidth, height =0.16\textwidth]{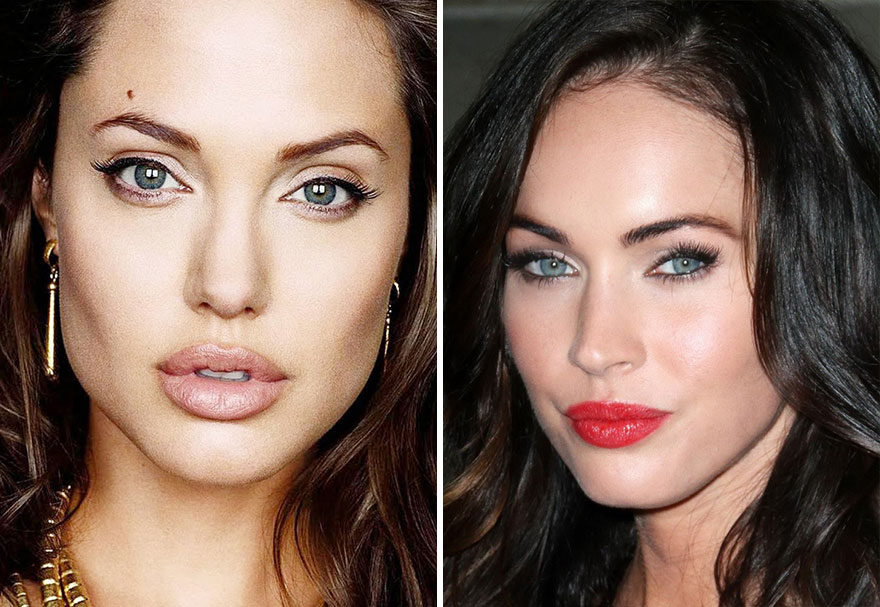}
    }
    \subfloat[]
    {
      \includegraphics[width=0.16\textwidth, height =0.16\textwidth]{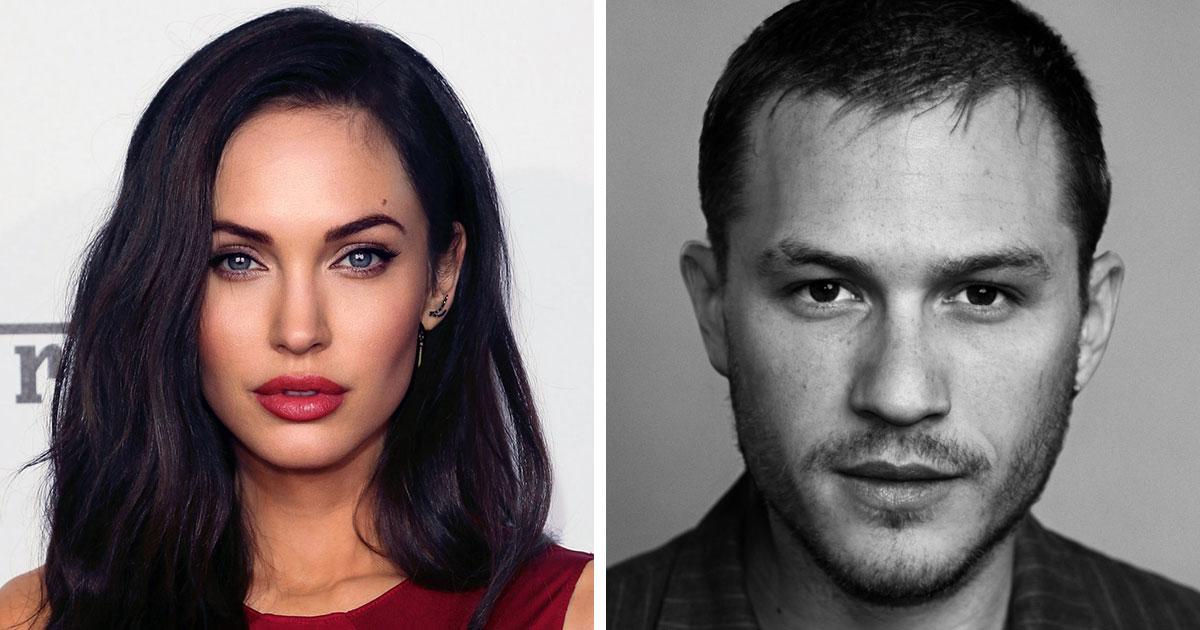}
    }
    \subfloat[]
    {
      \includegraphics[width=0.16\textwidth, height =0.16\textwidth]{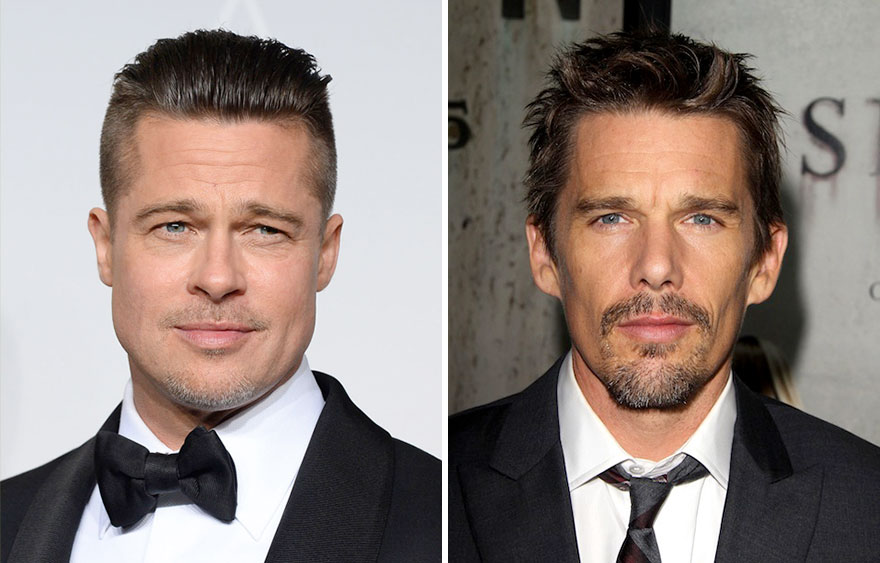}
    }
    \caption{Real-world examples of combined images from Google search.}
        \label{figures-combined-google-search}
\end{figure}

\end{document}